\documentclass[a4paper,fleqn,usenatbib]{mnras}

\usepackage[T1]{fontenc}
\usepackage{ae,aecompl}

\usepackage{graphicx}	
\usepackage{amsmath}	
\usepackage{amssymb}	
\usepackage{color}
\definecolor{ao(english)}{rgb}{0.0, 0.5, 0.0}
\usepackage{ulem}

\title[X-rays from stellar bow shocks]{X-ray study of bow shocks in runaway stars
\thanks{Based on observations with XMM-Newton, an ESA Science Mission with instruments and contributions directly funded by ESA Member states and the USA (NASA).}}
\author[M. De Becker et al.]{M. De Becker$^{1}$,\thanks{E-mail:
debecker@astro.ulg.ac.be}  M.V. del Valle$^{2}$, G.E. Romero$^{3,4}$, C.S. Peri$^{3,4}$, P. Benaglia$^{3,4}$\\
$^{1}$Space sciences, Technologies and Astrophysics Research (STAR) Institute, University of Li\`ege, Quartier Agora, 19c, All\'ee du 6 Ao\^ut,\\ B5c, B-4000 Sart Tilman, Belgium\\
$^{2}$Institute of Physics and Astronomy, University of Potsdam, 14476 Potsdam, Germany\\
$^{3}$Instituto Argentino de Radioastronom\'{\i}a (CONICET;CICPBA), C.C. No 5, 1894, Villa Elisa, Argentina\\
$^{4}$Facultad de Ciencias Astron\'{o}micas y Geof\'{\i}sicas, UNLP, Paseo del Bosque s/n, 1900, La Plata, Argentina
}

\pubyear{2017}
\graphicspath{./figs/}
\begin{document}
\label{firstpage}
\pagerange{\pageref{firstpage}--\pageref{lastpage}}
\maketitle

\begin{abstract}
{
Massive runaway stars produce bow shocks through the interaction of their winds with the interstellar medium, with the prospect for particle acceleration by the shocks. These objects are consequently candidates for non-thermal emission. Our aim is to investigate the X-ray emission from these sources. We observed with XMM-Newton a sample of 5 bow shock runaways, which constitutes a significant improvement of the sample of bow shock runaways studied in X-rays so far. A careful analysis of the data did not reveal any X-ray emission related to the bow shocks. However, X-ray emission from the stars is detected, in agreement with the expected thermal emission from stellar winds. On the basis of background measurements we derive conservative upper limits between 0.3 and 10 keV on the bow shocks emission. Using a simple radiation model, these limits together with radio upper limits allow us to constrain some of the main physical quantities involved in the non-thermal emission processes, such as the magnetic field strength and the amount of incident infrared photons. The reasons likely responsible for the non-detection of non-thermal radiation are discussed. Finally, using energy budget arguments, we investigate the detectability of inverse Compton X-rays in a more extended sample of catalogued runaway star bow shocks. From our analysis we conclude that a clear identification of non-thermal X-rays from massive runaway bow shocks requires one order of magnitude (or higher) sensitivity improvement with respect to present observatories.    
}

\end{abstract}




\section{Introduction}
Most massive stars, mainly OB-type stars along with Wolf-Rayet stars (their evolved counterparts) are known to be part of open clusters, where they are believed to be formed. However, some stars -- the so-called runaways or runaway stars -- are characterized by large peculiar space velocities and seem to have been ejected from their birth place. Mainly, two scenarios have been proposed to explain the peculiar kinematics of runaway stars. On the one hand, one may consider a supernova explosion in a close binary (or higher multiplicity) system, resulting in the ejection of the secondary \citep{zwicky}. On the other hand, the ejection of the star can also be the result of dynamical interactions in a dense open cluster \citep{leonardduncan}.

\begin{table*}
\caption{Stellar and bow shock parameters. The bolometric luminosities for O-stars are the typical values given by \citet{martins}. For B-stars we considered the values given by \citet{hns2010}. All other parameters are taken from the E-BOSS catalogue (see the text for further description). \label{param}}
\begin{center}
\begin{tabular}{c c c c c c c c c c c c c c}
\hline
\multicolumn{3}{c}{Target} & & \multicolumn{2}{c}{General} & & \multicolumn{2}{c}{Wind} &  & \multicolumn{4}{c}{Bow shock} \\
\cline{1-3}\cline{5-6}\cline{8-9}\cline{11-14}
\vspace*{-0.2cm}\\
ID  & Altern. ID & Spec. type & & d & L$_\mathrm{bol}$ & & ${\dot M}$ & $V_\infty$ &  & $l$ & $w$ & $R$ & $R$ \\
 &  &  &  & (pc) & (erg\,s$^{-1}$) & & (M$_\odot$\,yr$^{-1}$) & (km\,s$^{-1}$) &  & (pc) & (pc) & (pc) & (')\\
\hline
\vspace*{-0.2cm}\\
HIP\,16518 & HD\,21856 & B1V & & 650 & 1.21\,$\times$\,10$^{38}$ & & 6.0\,$\times$\,10$^{-9}$ & 500 & & 0.76 & 0.19 & 0.13 & 0.7 \\
HIP\,34536 & HD\,54662 & O6.5V((f)) + O9V & & 1293 & 8.22\,$\times$\,10$^{38}$ & & 1.9\,$\times$\,10$^{-7}$ & 2456 & & 4.51 & 1.13 & 1.50 & 4 \\
HIP\,77391 & HD\,329905 & O9I & & 800 & 1.35\,$\times$\,10$^{39}$ & & 2.5\,$\times$\,10$^{-7}$ & 1990 & & 0.93 & 0.23 & 0.23 & 1 \\
HIP\,78401 & HD\,143275 & B0.2IVe & & 224 & 2.58\,$\times$\,10$^{38}$ & & 1.4\,$\times$\,10$^{-7}$ & 1100 & & 1.63 & 0.13 & 0.39 & 6 \\
HIP\,97796 & HD\,188001 & O7.5I & & 2200 & 1.70\,$\times$\,10$^{39}$ & & 5.0\,$\times$\,10$^{-7}$ & 1980 & & 8.32 & 1.60 & 3.84 & 6 \\
\vspace*{-0.2cm}\\
\hline
\end{tabular}
\end{center}
\end{table*}

Among these systems, several tens turn out to produce strong shocks, through the interaction of their stellar winds with the interstellar material, as described in the Extensive stellar BOw Shock Survey (E-BOSS) catalogue \citep{ebossI,ebossII}. The existence of these shocks -- often revealed in the infrared -- calls upon the question of particle acceleration through the Diffusive Shock Acceleration mechanism (DSA) \citep{fermi,bell1978,drury1983}. This process is indeed operating efficiently in other astrophysical environments presenting hydrodynamic shocks such as supernova remnants \citep{romerosnr,VinkSNRutrecht} and particle-accelerating colliding-wind binaries \citep{PD140art,debeckerreview,catapacwb}. Predictions about the non-thermal activity of runaways with bow shocks were made by \citet{ntbowshock}. The action of particle acceleration was indeed confirmed indirectly through the detection of synchrotron radiation in the radio domain for BD\,+43$^\circ$3654 \citep{benagliarunaway}. In X-rays, a putative detection of non-thermal emission was reported in the case of AE\,Aurigae \citep{aeaurigaexmm}. Such an emission would be interpreted in terms of inverse Compton (IC) scattering of infrared photons produced by the interstellar dust heated by the stellar photons, in the presence of a population of relativistic electrons accelerated through DSA \citep{ntbowshock}. More recently, the X-ray observation of a few massive runaways (including BD\,+43$^\circ$3654) did not reveal any non-thermal emission at the expected position of their bow shocks \citep{toala2016,toala2017}. In $\gamma$-rays, the scenario where the runaway HD\,195592 could be at the origin of the Fermi source 2FGL\,J2030.7\,+4417 was investigated by \citet{DRD2013}, though this $\gamma$-ray source is probably predominantly associated to a pulsar \citep{abdo2013}. The investigation for Fermi counterparts of a larger sample of bow shock runaways performed by \citet{schulzbsr} failed to detect any of them in $\gamma$-rays. More recently, the systematic search for higher energy gamma-ray emission from all members of the E-BOSS catalogue using the  H.E.S.S. (High Energy Stereoscopic System) observatory did not lead to any detection neither \citep{hessbs}.

So far, the detection of high energy non-thermal radiation from bow shock runaways remains thus elusive, justifying a dedicated effort from the observational point of view to clarify this issue. To do so, we obtained observation time with the {\it XMM-Newton} satellite to observe in X-rays a sample of 5 OB-type runaways with bow shocks. The main objective of this study is to investigate the potential non-thermal emission from these bow shocks, with the prospects to feed recent models relevant for this class of objects with actual measurements. Considering the low number of similar observations in the past, our study provides a significant improvement for the sample of bow shock runaways investigated in X-rays.

The paper is organized as follows. The selection criteria of the sample and the main basic properties of our targets are presented in Sect.\,\ref{sample}. Section\,\ref{obser} describes the observations and the general data processing. A more detailed description of the specific results obtained for all targets and their discussion in the appropriate context are provided, respectively, in Sect.\,\ref{sectresults} and Sect.\,\ref{disc}. Finally, the summary and conclusions are given in Sect.\,\ref{concl}.

\section{Description of the sample}\label{sample}
The targets were selected among the members of the first release of the E-BOSS catalogue \citep{ebossI}. Without any a priori idea of the potential non-thermal emission from the bow shocks of these objects, we scaled the X-ray emission level reported by \citet{aeaurigaexmm} for AE\,Aurigae taking into account two factors. First, the scaling took into account the distance to the potential target (closer targets would lead to a higher flux). Second, considering the source photons for IC scattering is IR radiation from dust particles, another scaling factor was used to account for the IR emission level of the bow shock with respect to that of AE\,Aurigae (the brighter the IR emission, the brighter the expected IC emission). We also rejected runaways whose bow shocks were located at angular distance smaller than 0.5 arcmin to reduce confusion with the expected X-ray emission from the stars. This procedure allowed to rank higher priority targets for observations with the {\it XMM-Newton} satellite. Finally, we restricted our selection procedure to objects that were never observed before with modern X-ray observatories.

Our sample consists of the 5 highest priority targets emerging from this selection procedure. Table\,\ref{param} summarizes the main properties of the targets that are relevant for the present study: d is the source distance and L$_\mathrm{bol}$ is its bolometric luminosity; $\mathrm{\dot M}$ and  V$_\infty$ are the wind mass-loss rate and terminal velocity, respectively\footnote{We caution that the wind parameters for B-type stars are certainly less well determined than those for O-type stars, with weaker winds lacking accurate predictions \citep[see e.g.][]{puls2008}. However, a change with respect to the adopted value would be speculative, with no substantial change in the results described in this paper.}. The bow shock parameters are the length $l$ and the stagnation radius $R$, i.e. the distance from the star to the midpoint of the bow shock structure (in both linear and angular units). The width $w$ can be viewed as the projected thickness of the bow shock. Most of these quantities will be used throughout Sections\,\ref{sectresults} and \ref{disc}. 

\section{Observations and data processing}\label{obser}

Observations were granted in the context of the 13th Announcement of Opportunity (AO13) of {\it XMM-Newton}, with programme ID 074366. A summary of the observations is given in Table\,\ref{journal}. The three EPIC cameras (MOS1, MOS2 and pn) were operated in the Full Frame mode with the medium filter to reject optical light. The aim point of all observations was set to the position of the stars. Data processing was performed using the Science Analysis Software (SAS) v.15.0.0 on the basis of the Observation Data Files (ODF) provided by the European Space Agency (ESA), using the calibration files (CCF) available in May 2016. We adopted the so-called standard screening criteria, namely pattern $\leq$\,12 for MOS and pattern $\leq$\,4 for pn. For pn data, we also adopted the adequate screening criteria to get rid of bright CCD borders and obtain clean images. The journal of observations for the 5 targets is shown in Table\,\ref{journal}. In each case, the difference between the performed duration (cols.\,4 and 5) and the effective exposure time (cols.\,6 and 7) is mainly explained by the rejection of time intervals contaminated by soft proton flares \citep{Lumb2002}. To do so, we extracted light curves (100\,s time bins) made of events with Pulse height Invariant (PI) larger than 10000, and we adopted various rejection thresholds (expressed in counts) depending on the observations and on the instrument (MOS or pn), as specified in columns 8 and 9 in Table\,\ref{journal}. In these columns, '--' means that no soft proton flaring activity required any time filtering.

\begin{table*}
\caption{Journal of observations. \label{journal}}
\begin{center}
\begin{tabular}{l l c c c c c c c c c}
\hline
Target & Observ. Id. & Date & \multicolumn{2}{c}{Performed duration (s)} & & \multicolumn{2}{c}{Eff. exp. time (s)} &  & \multicolumn{2}{c}{Rejection threshold (cnt)}\\
\cline{4-5}\cline{7-8}\cline{10-11}
  &  &  & MOS & pn & & MOS & pn &  & MOS & pn\\
 (1) & (2) & (3) & (4) & (5) & & (6) & (7) & & (8) & (9) \\
\hline
\vspace*{-0.2cm}\\
HIP\,16518 & 0743660301 & 3 March 2015 & 36700 & 35000 & & 26860 & 22350 & & 20 & 150 \\
HIP\,34536 & 0743660501 & 1 October 2014 & 32600 & 30900 & & 30320 & 23920 & & 15 & 150 \\
HIP\,77391 & 0743660401 & 16 March 2015 & 39400 & 37700 & & 35530 & 26140 & & 15 & 150 \\
HIP\,78401 & 0743660101 & 7 March 2015  & 38700 & 37000 & & 30449 & 18080 & & 20 & 200 \\
HIP\,97796 & 0743660201 & 14 October 2014 & 31700 & 30000 & & 31340 & 26760 & & -- & -- \\
\vspace*{-0.2cm}\\
\hline
\end{tabular}
\end{center}
\end{table*}

Images were obtained for all EPIC cameras in the 0.3--10.0\,keV energy range. A first inspection of the images revealed point sources at the position of all stars, except for HIP\,77391 only weakly present on the pn image. Beside the stars, other point sources are also present in all fields of view. For all targets bright enough in X-rays for spectral analysis, circular spatial filters were used to extract source and background spectra. Details about their location and size are given specifically for all objects in Section\,\ref{specanal}. Response matrices and ancillary response files were computed using the {\tt rmfgen} and {\tt arfgen} metatasks implemented in the SAS. Spectra were grouped to get at least 5 events per energy bin to avoid the issue of too low number statistics.

\section{Results}\label{sectresults}

\subsection{Modelling of X-ray spectra}\label{models}
When the extraction of spectra of detected sources was relevant, we made use of the XSPEC software (v.12.8.2, \citealt{xspec1996,xspec2001}) for the spectral analysis. The adopted goodness-of-fit indicator was the $\chi^2_\nu$ value using the Churazov weighting adequate for low count numbers \citep{churazov}. Error bars on all model parameters were computed on the basis of the {\tt error} command within XSPEC. This offers the advantage to adequately estimate confidence intervals through a systematic exploration of the parameter space. This allows moreover to refine the spectral fitting and it favors the convergence to a best-fit model. Solar abundances were assumed throughout this analysis.

Spectra were modelled using composite models, whose components have been selected depending on the expected processes likely to be at work in the investigated objects. The expected thermal X-ray emission from the stellar winds is adequately represented by an optically thin thermal plasma emission component. In the standard model for X-ray emission from massive star winds, hydrodynamic shocks occur in stellar winds because of the line-driving instability, resulting in the heating of the plasma up to temperatures of a few MK \citep{FeldX}. To account for this process, we made use of the {\tt apec} model. One of the main parameters is the plasma temperature kT, expressed in keV. The other important parameter is the normalization (N), directly related to the emission measure of the emitting plasma\footnote{See the XSPEC User's Handbook for details, {\tt https://heasarc.gsfc.nasa.gov/xanadu/xspec/manual/manual.html}}). The non-thermal X-ray emission expected to be produced close to the bow shock can be represented by a power law emission component, i.e. the {\tt power} component within XSPEC. This model is defined by a photon index $\Gamma$, and by a normalization parameter translating the flux emitted by the non-thermal component. The photon index is directly related to the electron index characterizing the power law relativistic electron distribution. In the absence of direct measurement of $\Gamma$, a value of 1.5 will be assumed. This value is anticipated for inverse Compton scattering due to relativistic electrons (p = 2) accelerated by DSA in the presence of standard strong shocks \citep{BG1970,ntbowshock}.

The absorption by the interstellar material was represented by a {\tt wabs} photoelectric absorption model, whose unique parameter is the equivalent hydrogen column (expressed in H equivalent atom numbers per cm$^2$). This parameter was determined assuming ${N_\mathrm{H,ISM}}$ = $5.8\,\times\,10^{21}\,\times\,E(B - V)$\,cm$^{-2}$ \citep{Boh}, with a colour excess determined from an observed $(B - V)$ based on SIMBAD magnitudes and an intrinsic colour $(B - V)_\circ$ based on the relation given by \citet{MB1981}. For all targets, the adopted values for ${N_\mathrm{H,ISM}}$ are given in Table\,\ref{nh}. For the emission from the stars (and not from the bow shocks), a second {\tt wabs} absorption component was used to account for the expected absorption by the stellar wind material. For a detailed discussion on the impact of stellar wind absorption on massive star X-ray spectra, we refer to \citet{Leut2010}. This second absorption component turned out to be relevant especially for the case of O-type stars, whose wind are strong enough to produce a significant absorption. The relevant parameter will be referred to as ${N_\mathrm{H,WIND}}$ in Sect.\,\ref{specanal}. We clarify that this second absorption component does make sense only for the X-ray emission from the embedded shocks in the stellar winds, and not for the putative emission from the extended bow-shocks.

\begin{table}
\caption{Adopted values for the interstellar absorption column $\mathrm{N_{H,ISM}}$. \label{nh}}
\begin{center}
\begin{tabular}{l c c c c}
\hline
Target & B & V & $E(B - V)_\circ$ & ${N_\mathrm{H,ISM}}$ \\
 & & & & (cm$^{-2}$) \\
\hline
\vspace*{-0.2cm}\\
HIP\,16518 & 5.82 & 5.90 & --0.30 & 0.13\,$\times$\,10$^{22}$ \\
HIP\,34536 & 6.24 & 6.21 & --0.31 & 0.20\,$\times$\,10$^{22}$ \\
HIP\,77391 & 10.94 & 10.45 & --0.30 & 0.47\,$\times$\,10$^{22}$ \\
HIP\,78401 & 2.20 & 2.32 & --0.30 & 0.11\,$\times$\,10$^{22}$ \\
HIP\,97796 & 6.24 & 6.23 & --0.31 & 0.19\,$\times$\,10$^{22}$ \\
\vspace*{-0.2cm}\\
\hline
\end{tabular}
\end{center}
\end{table}

\subsection{Analysis of X-ray images and spectra}\label{specanal}

Our main objective is to investigate the potential X-ray emission associated to bow shocks. To do so, a careful analysis of images is first required. Beside the search for a diffuse emission extending over the bow shock region, one should also consider the presence of point-like sources likely attributable to a bow shock emission that peaks above the background level at a given position. As a result, the correlation of point-like sources with infrared and visible catalogues is needed to reject independent X-ray emitters in the vicinity of the bow shock. Following this approach, only significant X-ray sources with no identified counterparts at other wavelengths could be relevant for the purpose of this study. Apart from our investigation of bow shocks, a spectral analysis of the stellar emissions is also preformed using the tools described in Sect.\,\ref{models}.

\subsubsection{HIP\,16518}\label{16518}
The runaway appears as a weak point source in the EPIC field of view. The X-ray count rate between 0.3 and 10.0\,keV is 0.004\,$\pm$\,0.001, 0.003\,$\pm$\,0.001 and 0.012\,$\pm$\,0.001\,cnt\,s$^{-1}$, respectively for MOS1, MOS2 and pn. The infrared image in the E-BOSS catalogue shows that the bow shock is rather close to the star, with a stagnation radius $R$ of about 0.7\,' (see Table\,\ref{param}). The pn image presents a slight apparent X-ray excess with respect to the surrounding background to the North-West direction, corresponding to the pointing direction of the runaway velocity vector. The correlation of its position with the 2-Micron All-Sky Survey (2MASS) point source catalogue \citep{2mass} and with the Guide Star Catalogue (2.3.2, STScI\footnote{In this paper, GSC counterparts appear as 10 digit identifiers.}) did not reveal any known counterpart within a radius of several arcsec. This motivates to consider that the X-ray excess -- if real -- could potentially be associated to the bow shock. Even though it is marginally detected, it will deserve a specific discussion below.

\begin{figure*}
\begin{center}
\includegraphics[width=16cm]{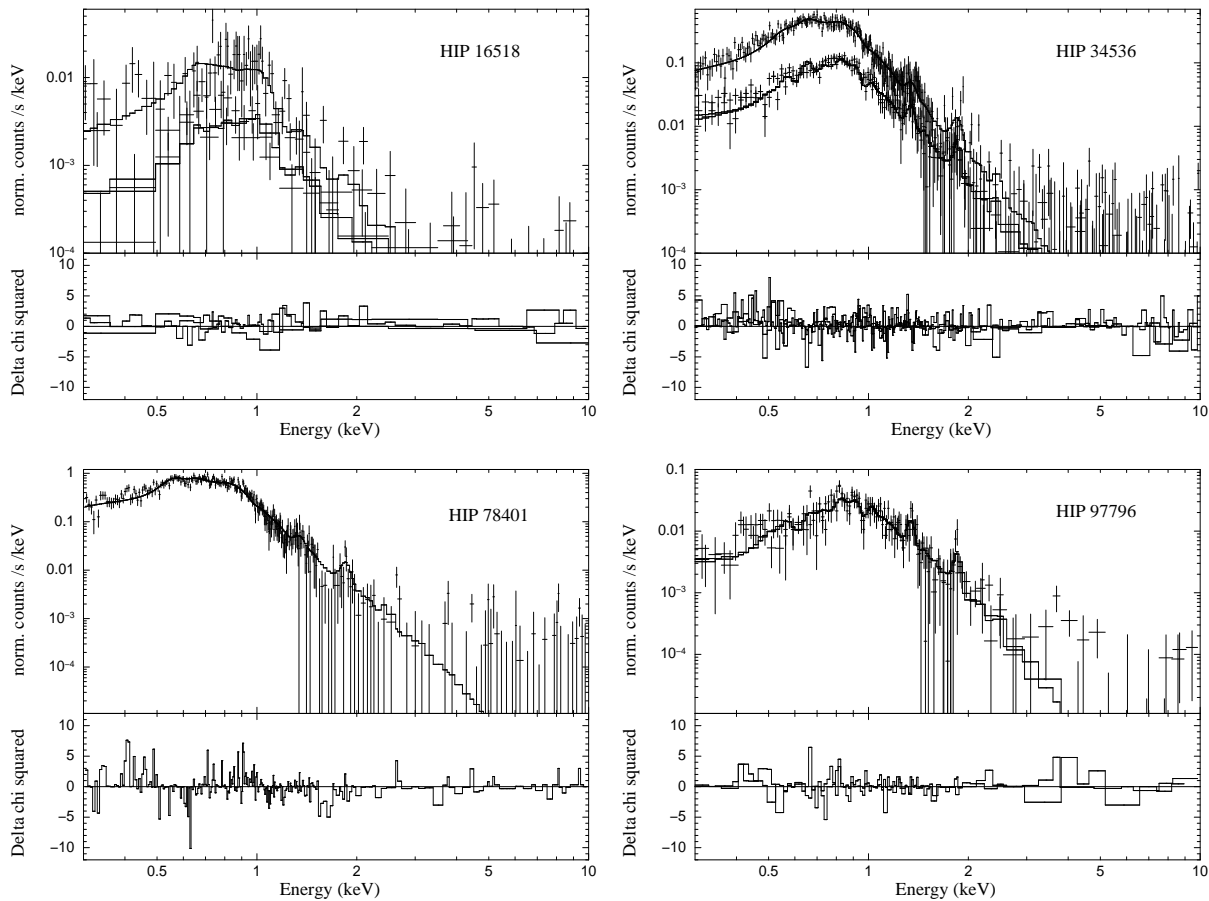}
\caption{EPIC spectra and their best-fit models (see Table\,\ref{fitstars}) between 0.3 and 10.0\,keV for HIP\,16518, HIP\,34536, HIP\,78401 and HIP\,97796.\label{spec}}
\end{center}
\end{figure*}

The spectrum of the star was extracted for all EPIC cameras from a circular region centred on the star coordinates, with a 30\,'' radius. The background was extracted within a circular region with equal size as the source region on the same detector (at [3:32:53.8; +35:29:31.4] for MOS and [3:32:31.6; +35:27:06.3] for pn), away from the expected position of the bow shock. The count rate of the source did not allow to confront spectra to sophisticated models including several components. We thus focus on the results obtained from a model with one absorption component and one thermal emission component (e.g., {\it wabs*apec}) fitted simultaneously to the three EPIC spectra (see Fig.\,\ref{spec}). The absorption column was fixed to the interstellar value (see Table\,\ref{nh}). In the context of this thermal model, the plasma temperature is of about 6-7\,MK. This range of temperature is in agreement with the typical values expected from intrinsic shocks in individual stellar winds.

\begin{figure}
\begin{center}
\includegraphics[width=8cm]{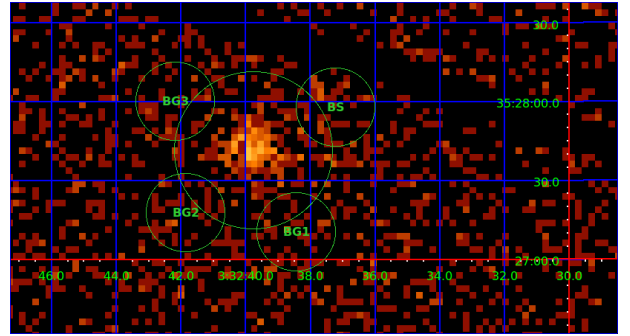}
\caption{EPIC pn image of the vicinity of HIP\,16518 showing the source region of the star (large circle), along with the extraction regions for the marginal excess coincident with the bow shock (BS) and the background regions (BG1, BG2 and BG3) used for the analysis.  \label{bs16518}}
\end{center}
\end{figure}

\begin{table}
\caption{Count numbers in the X-ray excess close to HIP\,16518 and in the nearby background. The value marked with a * is the net count number (i.e. corrected for the mean background count number $\mathrm{<C_{BG}>}$) associated to the slight excess.\label{excessres}}
\begin{center}
\begin{tabular}{l c c c c}
\hline
Region & RA & DEC & C & $\mathrm{C - <C_{BG}>}$  \\
 &  &  & (cnt) & (cnt) \\
\hline
\vspace*{-0.2cm}\\
BS & 3:32:37:21 & +35:27:57.53 & 38 & 8* \\
BG1 & 3:32:38.43 & +35:27:10.38 & 33 & 3 \\
BG2 & 3:32:41.84 & +35:27:17.64 & 26 & 4 \\
BG3 & 3:32:42.17 & +35:27:59.84 & 30 & 0 \\
\vspace*{-0.2cm}\\
\hline
\end{tabular}
\end{center}
\end{table}

The slight X-ray excess coincident with the position of the bow shock has been measured using circular extraction regions as illustrated in Fig.\,\ref{bs16518}. We extracted events related to this emission excess using a 15\,'' radius circular spatial filter, with rejection of the events falling in the extraction region of the runaway (see above). We have to be conscious that the excess appears in the wings of the Point Spread Function (PSF) of the stellar source. In order to correct for this, we estimated the background level at the position of the X-ray excess through the extraction of events in identical regions equidistant from the star (regions BG1, BG2 and BG3 in Fig.\,\ref{bs16518}). These background regions are expected to be affected at the same level by the wings of the PSF of the stellar source than the excess extraction region. These extraction regions were used to measure the count numbers in the excess and in the background (mean of the 3 regions), respectively. The results are summarized in Table\,\ref{excessres}. These numbers lead to a standard deviation for the background level of about 3 counts. With a count number of about 8 in the excess region, the numbers suggest a detection slightly below the 3\,$\sigma$ level. We will therefore consider it is marginal.

Assuming the slight excess is real and behaves like a point source, one should correct for the encircled energy fraction of a 15\,'' region, i.e. about 75\,\%. An additional correction should be provided to take into account that the extraction region is not circular, as it is truncated by the rejection of the star extraction region (roughly, 10\,\% loss). The corrected count number for the excess is thus of the order of 12. Dividing by the effective exposure time of the pn observation, one obtains a count rate of about 5.4\,$\times$\,10$^{-4}$\,cnt\,s$^{-1}$ for the slight X-ray excess coincident with the bow shock of HIP\,16518.

The WebPIMMS tool\footnote{https://heasarc.gsfc.nasa.gov/cgi-bin/Tools/w3pimms/w3pimms.pl}, developed and hosted by the High Energy division at the National Aeronautics and Space Administration (NASA), was used to convert this count rate into a physical flux. Assuming an inverse Compton scattering emission process, we considered an absorbed (by the interstellar medium) power law emission spectrum. Using the ${N_\mathrm{H,ISM}}$ value given in Table\,\ref{nh} and a photon index equal to 1.5, we derived an X-ray flux of 2.3\,$\times$\,10$^{-15}$\,erg\,cm$^{-2}$\,s$^{-1}$ between 0.3 and 10.0\,keV. In the same energy band, the flux corrected for the interstellar absorption is 3.52\,$\times$\,10$^{-15}$\,erg\,cm$^{-2}$\,s$^{-1}$. With a distance of 650\,pc, we obtain a potential inverse Compton luminosity for the slight X-ray excess coincident with the bow shock $\mathrm{L_{X,excess} = 1.4\,\times\,10^{29}}$\,erg\,s$^{-1}$.

\begin{table*}
\caption{Best-fit parameters for the runaway stars using thermal models (in the 0.3--10\,keV energy band). The confidence intervals at the 90\,\% level are specified for all parameters and for the observed flux.\label{fitstars}}
\begin{center}
\begin{tabular}{l c c c c c c c c}
\hline
Target & Instr. & $N_\mathrm{H,WIND}$ & kT$_1$ & N$_1$ & kT$_2$ & N$_2$ & $\chi^2_\nu$ (d.o.f.) & f$_{\rm obs}$ \\
 &  & (10$^{22}$\,cm$^{-2}$) & (keV) & (cm$^{-5}$) & (keV) & (\,cm$^{-5}$) &  & (erg\,cm$^{-2}$\,s$^{-1}$) \\
\hline
\vspace*{-0.2cm}\\
HIP\,16518 & EPIC & -- & 0.76$_{0.71}^{0.83}$ & 7.3$_{6.5}^{8.1}$\,$\times$\,10$^{-6}$ & -- & -- & 1.18 (130) & 1.52$_{1.37}^{1.61}$\,$\times$\,10$^{-15}$\\
\hline
\vspace*{-0.2cm}\\
HIP\,34536 & EPIC & 0.19$_{0.16}^{0.23}$ & 0.13$_{0.10}^{0.15}$ & 3.4$_{2.0}^{9.6}$\,$\times$\,10$^{-3}$ & 0.28$_{0.25}^{0.31}$ & 1.0$_{0.7}^{1.3}$\,$\times$\,10$^{-3}$ & 1.50 (597) & 4.25$_{2.86}^{4.27}$\,$\times$\,10$^{-13}$\\
\vspace*{-0.2cm}\\
 & EPIC & 0.07$_{0.03}^{0.09}$ & 0.20$_{0.19}^{0.21}$ & 1.0$_{0.7}^{1.2}$\,$\times$\,10$^{-3}$ & 0.58$_{0.56}^{0.61}$ & 1.5$_{1.3}^{1.5}$\,$\times$\,10$^{-4}$ & 1.51 (597) & 4.29$_{4.17}^{4.34}$\,$\times$\,10$^{-13}$\\
\hline
\vspace*{-0.2cm}\\
HIP\,78401 & MOS & -- & 0.21$_{0.20}^{0.23}$ & 5.3$_{5.0}^{5.7}$\,$\times$\,10$^{-4}$ & 0.61$_{0.58}^{0.65}$ & 1.2$_{0.9}^{1.4}$\,$\times$\,10$^{-4}$ & 1.77 (233) & 6.58$_{6.43}^{6.76}$\,$\times$\,10$^{-13}$\\
\vspace*{-0.2cm}\\
 & pn & -- & 0.19$_{0.18}^{0.20}$ & 6.6$_{6.3}^{6.9}$\,$\times$\,10$^{-4}$ & 0.59$_{0.57}^{0.63}$ & 1.5$_{1.3}^{1.7}$\,$\times$\,10$^{-4}$ & 1.13 (279) & 7.49$_{7.36}^{7.64}$\,$\times$\,10$^{-13}$\\
\hline
\vspace*{-0.2cm}\\
HIP\,97796 & MOS & 0.46$_{0.41}^{0.51}$ & 0.24$_{0.23}^{0.26}$ & 1.5$_{1.0}^{2.1}$\,$\times$\,10$^{-3}$ & -- & -- & 1.30 (124) & 1.27$_{1.10}^{1.32}$\,$\times$\,10$^{-13}$\\
\vspace*{-0.2cm}\\
 & pn & 0.52$_{0.49}^{0.54}$ & 0.20$_{0.19}^{0.21}$ & 3.7$_{3.1}^{4.6}$\,$\times$\,10$^{-3}$ & -- & -- & 1.81 (350) & 1.34$_{1.11}^{1.37}$\,$\times$\,10$^{-13}$\\
\vspace*{-0.2cm}\\
 & MOS & 0.44$_{0.34}^{0.56}$ & 0.12$_{0.11}^{0.15}$ & 8.8$_{2.4}^{44.0}$\,$\times$\,10$^{-3}$ & 0.48$_{0.37}^{0.59}$ & 2.3$_{1.3}^{5.0}$\,$\times$\,10$^{-4}$ & 0.93 (181) & 1.42$_{0.77}^{1.44}$\,$\times$\,10$^{-13}$\\
\vspace*{-0.2cm}\\
 & pn & 0.13$_{0.07}^{0.17}$ & 0.06$_{0.04}^{0.07}$ & 8.5$_{1.2}^{10.8}$\,$\times$\,10$^{-2}$ & 0.63$_{0.59}^{0.66}$ & 1.1$_{0.9}^{1.4}$\,$\times$\,10$^{-4}$ & 1.19 (257) & 1.62$_{1.26}^{1.69}$\,$\times$\,10$^{-13}$\\
\vspace*{-0.2cm}\\
\hline
\end{tabular}
\end{center}
\end{table*}

\subsubsection{HIP\,34536}\label{34536}
The star appears as a moderately bright star in X-rays, with count rates of 0.065\,$\pm$\,0.002, 0.064\,$\pm$\,0.002 and 0.268\,$\pm$\,0.038\,cnt\,s$^{-1}$, respectively for MOS1, MOS2 and pn between 0.3 and 10.0\,keV. The source spectra were extracted in circular regions with 60\,'' radius centered at the position of the runaway. The background regions for MOS and pn were circles with 60\,'' radius at [7:09:07.9; --10:22:36.5] and [7:09:17.4; --10:23:12.8], respectively. The spectral modelling using XSPEC led to satisfactory results with a two-temperature model absorbed by both interstellar and wind columns (i.e. {\tt wabs*wabs*(apec+apec)}). Fit of individual data sets and the simultaneous fit of the three EPIC spectra (see Fig.\,\ref{spec}) gave consistent results. Only the latter parameter values are quoted in Table\,\ref{fitstars}. We note that two solutions appeared to be quite similar in terms of goodness-of-fit and physical relevance, with different temperature distributions. Nonetheless both solutions point to a rather soft spectrum with plasma temperatures of a few MK at most.

\begin{figure}
\begin{center}
\includegraphics[width=8cm]{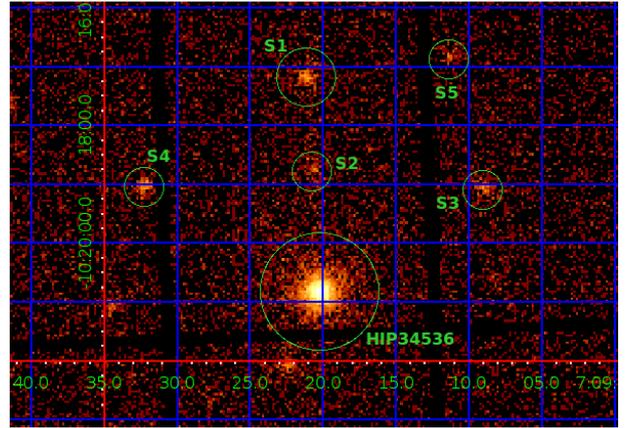}
\caption{EPIC pn image of the vicinity of HIP\,34536 showing the extraction region of the star (large circle), along with the position of the point sources S1 to S5 (see text).  \label{reg34536}}
\end{center}
\end{figure}

The region where the bow shock is located presents a few point sources (see Fig.\,\ref{reg34536}). Their positions were correlated with the 2MASS point source catalogue and with the GSC catalogue. It appears that most of these sources possess very likely counterparts: S2 at [7:09:20.54; --10:18:41.27] is at 1.2\,'' of 2MASS\,07092061-1018408, S3 at [7:09:09.02; --10:19:06.26] is at 4.5\,'' of 2MASS\,07090904-1019017 and S3TJ000659, S4 at [7:09:32.21; --10:19,02.59] is at 3.2\,'' of 2MASS\,07093222-1018592 and S3TJ027311, an S5 at [7:09:11.25; --10:16,48.89] is at 1.1\,'' of S3TH016574. This rejects with a high level of certainty the possibility that these point-like sources may be physically related to the bow shock. However, EPIC images reveal the presence of a source (S1) at coordinates [7:09:21.10; --10:17:09.53], located at a distance of about 3.6\,'' from the runaway. It is thus quite close to the stagnation radius of the bow shock (see Table\,\ref{sample}). We did not find any 2MASS or GSC counterpart at an angular distance less than 14\,'' for this X-ray source. This allows to consider the scenario where this unidentified X-ray source is related to the bow shock.

The presence of the moderately X-ray bright runaway introduces out-of-time event features affecting the position of S1. We thus corrected the event list for out-of-time events and we extracted a spectrum using a circular region of 20'' radius. The pn count rate is 0.009\,$\pm$\,0.001\,cnt\,s$^{-1}$, and MOS1 and MOS2 count rates are 0.003\,$\pm$\,0.001 and 0.004\,$\pm$\,0.001\,cnt\,s$^{-1}$, respectively. These values were estimated on the basis of background corrected spectra readable in XSPEC. Despite the low quality of spectra, we tentatively fitted models simultaneously to the three EPIC spectra. A {\tt wabs*wabs*power} model, with fixed interstellar absorption, gives a photon index of 1.64$_{1.27}^{2.09}$ with a measured flux of 6.4$_{5.5}^{6.9}$\,$\times$\,10$^{-14}$\,erg\,cm$^{-2}$\,s$^{-1}$. We caution however that a significant additional absorption column ($N_\mathrm{H}$ = 0.54$_{0.22}^{0.97}$\,$\times$\,10$^{22}$\,cm$^{-2}$) is needed to model the spectrum ($\chi^2_\nu$ of 1.04, for 132 d.o.f.). This may indicate a longer distance line-of-sight object with no physical relation with the bow shock. A confrontation of the spectra to a thermal model ({\tt wabs*wabs*apec}) led to a physically irrelevant result with an exceedingly high temperature (lower limit of about 4 keV) and a local absorbing column $N_\mathrm{H}$ = 0.43$_{0.16}^{0.87}$\,$\times$\,10$^{22}$\,cm$^{-2}$. These values are mentioned for the sake of completeness but we caution that the quality of the spectra introduces severe doubts on their reliability.

\subsubsection{HIP\,77391}\label{77391}
The B-star is undetected in MOS data and appears as a very faint point source in the pn field of view, with a count rate of 0.007\,$\pm$\,0.001\,cnt\,s$^{-1}$. This low level X-ray emission may be in part attributed to a faint and soft intrinsic emission from the weak stellar wind, along with a significant interstellar absorption toward that direction. This target is indeed characterized by the highest column density in the sample (see Table\,\ref{nh}). No relevant spectrum could therefore be extracted for the spectral analysis. Assuming an absorbed thermal emission model in the WebPIMMS tool, with a plasma temperature of 0.6 keV\footnote{This value is in agreement with the typical post-shock plasma temperature of presumably single star winds, with velocity of embedded shocks of a few 100\,km\,s$^{-1}$. Observational examples can be found for instance in \citet{debecker2013}, for O-type stars very likely to be single.}, we obtain a flux of $\sim$\,10$^{-14}$\,erg\,cm$^{-2}$\,s$^{-1}$ and an unabsorbed flux of $\sim$\,9\,$\times$\,10$^{-14}$\,erg\,cm$^{-2}$\,s$^{-1}$.

The careful inspection of the pn image reveals a marginal point-like source about 1 arcmin to the South of HIP\,77391. This direction and angular distance are coincident with the expected position of the bow shock. However, a correlation of this position with the 2MASS catalogue for point sources suggests this X-ray source should be associated to an independent object, with no physical relation with the bow shock. The infrared sources 2MASS\,15475227-4838578 and 2MASS\,15475263-4839035 lay indeed at angular distances of 4.2 and 4.8\,'' of the X-ray source, respectively. The same is true with the GSC catalogue with the source S8U6003194 at about 4.4\,''. We therefore do not detect any X-ray source likely attributable to the bow shock close to HIP\,77391. Upper limits on the putative X-ray emission from this bow shock are determined in Sect.\,\ref{upperlimits}.

\subsubsection{HIP\,78401}\label{78401}
The EPIC spectrum of HIP\,78401 was extracted from a circular region with a radius of 60\,'', excluding events in a 30'' radius circle associated to the nearby source at [16:00:17.14; -22:36:53.47]. The later source is certainly UCAC3 135-174588 (2MASS\,J16001730-2236504) referenced as a M4 star by \citet{LM2012}. The background spectrum was extracted in a circular region (radius 60\,'') at coordinates [16:00:35.01; --22:34:38.41] for MOS and [16:00:29.0; --22:34:04.97] for pn. The best fit was obtained with a thermal emission model with temperatures of about 2\,MK and 6\,MK. We note that even though globally the three individual EPIC data sets led to very similar results, the $\chi_\nu^2$ degraded significantly when the model was adjusted simultaneously to the three spectra. For this reason, MOS (see Fig.\,\ref{spec}) and pn results are presented separately in Table\,\ref{fitstars}. 

X-ray point sources positionally coincident with the bow shock have all identified counterparts in other wavebands: a source at [16:00:22.84; --22:43:54.90] is at about 4.1\,'' of 2MASS\,16002305-2243578 and S8ZL002730, a source at [16:00:16.66; --22:48:16.00] is at about 2.8\,'' of S8ZL046027, and a source at [15:59:57.36; --22:43:44.83] is at an angular distance of 1\,'' from S8ZL002780. Beside these point sources, no X-ray excess is detected close to the bow shock of HIP\,78401.

We note the presence of a diffuse X-ray source close to coordinates [16:00:07.5; --22:47:44.7], with no identified counterpart in the SIMBAD database. Even though this diffuse source is located in the appropriate direction with respect to the runaway, it is located too far away ($\sim$\,10\,arcmin) to be physically associated to the bow shock. The measured pn count rate is 0.013\,$\pm$\,0.002\,cnt\,s$^{-1}$, which is too faint for the extraction of a spectrum adequate for analysis. 

\subsubsection{HIP\,97796}\label{97796}

HIP\,97796 appears as a moderately bright source. The count rates are 0.023\,$\pm$\,0.001, 0.022\,$\pm$\,0.001 and 0.098\,$\pm$\,0.002\,cnt\,s$^{-1}$, respectively for MOS1, MOS2 and pn between 0.3 and 10.0\,keV. The source spectrum was extracted in a circular region of 45\,'' radius. The MOS and pn spectra were extracted using spatial filters with the same area centred at [19:52:06.65;+18:42:19.64] and [19:52:24.60;+18:31:09.70], respectively.  The spectrum is compatible with an optically thin thermal plasma emission model, with a temperature of the order of 2\,MK, or with two temperatures of about 1 and 6\,MK. Trials with a two-temperature model led apparently to a best fit but the relative error on the normalization parameter of the softer emission component is especially high. The results of the simultaneous fit to MOS spectra and individually to the pn spectrum are displayed in Table\,\ref{fitstars}. The best-fit of the two-temperature model to MOS spectra is shown in Fig.\,\ref
 {spec}.

Once again, no detectable diffuse emission coincident with the bow shock appears in the EPIC field. A few point sources are nonetheless present. Towards a direction close to that of the velocity vector of the runaway, a source at [19:52:15.28; +18:34:43.70] is coincident with 2MASS\,19521527+1834444 (at 0.7\,'') and N1TX000599 (at 0.6\,''). Slightly to the East, a point source at [19:52:28.29; +18:34:18.72] has counterparts at about 4\,'' (2MASS\,19522829+1834229 and N1TX065145). We note also the presence of a point source with no identified counterparts at other wavelengths at [19:52:23.54; +18:34:06.22]. We measured about 172 counts (corrected for the exposure map and for the average background measured in Sect.\,\ref{upperlimits}). This converts into a count rate of about 6.4\,$\times$\,10$^{-3}$\,cnt\,s$^{-1}$. Using the WebPIMMs tool, we assumed an absorbed power law spectrum expected to be adequate if this source is due to non-thermal emission associated to the bow shock. Between 0.3 and 10.0\,keV, one obtains a putative flux of 2.7\,$\times$\,10$^{-14}$\,erg\,cm$^{-2}$\,s$^{-1}$ and a flux corrected for interstellar absorption of 3.2\,$\times$\,10$^{-14}$\,erg\,cm$^{-2}$\,s$^{-1}$. We caution however that this point source is likely to be completely unrelated to the bow shock. In that case, this emission could probably be thermal, but we are lacking any indication about its typical plasma temperature. This prevents us to make any assessment on its nature.

\subsection{Upper limits on the non-thermal emission}\label{upperlimits}
We estimated upper limits on the count rate ($CR_\mathrm{max}$) potentially associated with the bow shock, on the basis of pn images only. MOS cameras collect indeed fewer events for a same exposure time and the pn results are thus expected to be more significant. We measured the number of counts ($C_\mathrm{i}$) in N regions devoid of obvious sources spread over the field of view, and we corrected those counts on the basis of the exposure map\footnote{The exposure map consists of a distribution across the EPIC field of view of the sensitivity of the camera. The sensitivity is decreasing from the centre to the borders of the field, with some gaps between CCDs of the mosaic.} computed with the SAS to scale them to an equivalent on-axis value ($C_\mathrm{i,corr}$),

$${C_\mathrm{i,corr} = C_i\,\frac{E_\mathrm{on}}{E_i}}$$
\noindent where ${E_\mathrm{on}}$ and ${E_\mathrm{i}}$ are the exposure map on-axis, and at the location of the estimation of the background level, respectively.

All extraction regions were arbitrarily selected as circles with a radius of 20\,''. We then computed the mean of the corrected count numbers, $${<C_\mathrm{corr}> = \frac{1}{N}\,\sum_\mathrm{i}^{N}C_\mathrm{i,corr}}$$

\noindent along with the deviation with respect to that mean,
$${\sigma_\mathrm{i} = |\,C_\mathrm{i,corr} - <C_\mathrm{corr}>\,|}$$
The typical fluctuation level of the background is thus estimated as the mean of the deviations:
$${\sigma = \frac{1}{N}\,\sum_\mathrm{i}^{N}\,\sigma_\mathrm{i}}$$
Finally, the division by the effective exposure time ${\Delta\,T_\mathrm{eff}}$ (col.\,7 in Table\,\ref{journal}) leads to a count rate. We adopted the assumption that the count numbers potentially produced by the bow shock should not be larger that three times the average fluctuation measured on the background (equivalent to 3$\sigma$ standard criterion). We thus establish the following expression for the estimator of the upper limit on the bow shock X-ray emission:
\begin{equation}\label{crmax}
{CR_\mathrm{max} = \frac{3\,\sigma}{\Delta\,T_\mathrm{eff}}}
\end{equation}

Adopting this procedure, we derived the upper limits quoted in Table\,\ref{ul} on the basis of a sample of $N = 5$ background regions. We caution that these upper limits depend intimately on the extraction radius, arbitrarily set to 20''. For this reason, and to ease any potential confrontation of these numbers to results obtained in future investigations, we also quote in Table\,\ref{ul} upper limits on the count rate densities (${CR^{d}_\mathrm{max}}$), expressed as the count rate per unit area (in cnt\,arcmin$^{-2}$\,s$^{-1}$).

\begin{table*}
\caption{Upper limits on the X-ray emission from the bow shocks measured in 20'' radius regions between 0.3 and 10.0\,keV. \label{ul}}
\begin{center}
\begin{tabular}{l c c c c c c}
\hline
Target & ${<C_\mathrm{corr}>}$ & $\sigma$ & ${CR_\mathrm{max}}$ & ${CR^{d}_\mathrm{max}}$ & $\mathrm{f_{IC,abs}}$ & $\mathrm{f_{IC}}$ \\
 & (cnt) & (cnt) & (cnt\,s$^{-1}$)& (cnt\,arcmin$^{-2}$\,s$^{-1}$) & (erg\,cm$^{-2}$\,s$^{-1}$) & (erg\,cm$^{-2}$\,s$^{-1}$) \\
\hline
\vspace*{-0.2cm}\\
HIP\,16518 & 58 & 9 & 1.2\,$\times$\,10$^{-3}$ & 3.4\,$\times$\,10$^{-3}$ & 1.3\,$\times$\,10$^{-14}$ & 1.5\,$\times$\,10$^{-14}$ \\
HIP\,34536 & 72 & 13 & 1.6\,$\times$\,10$^{-3}$ & 4.6\,$\times$\,10$^{-3}$ & 2.0\,$\times$\,10$^{-14}$ & 2.4\,$\times$\,10$^{-14}$ \\
HIP\,77391 & 156 & 31 & 3.6\,$\times$\,10$^{-3}$ & 10.3\,$\times$\,10$^{-3}$ & 5.5\,$\times$\,10$^{-14}$ & 7.2\,$\times$\,10$^{-14}$ \\
HIP\,78401 & 124 & 11 & 1.8\,$\times$\,10$^{-3}$ & 5.2\,$\times$\,10$^{-3}$ & 2.0\,$\times$\,10$^{-14}$ & 2.3\,$\times$\,10$^{-14}$ \\
HIP\,97796 & 82 & 4 & 4.5\,$\times$\,10$^{-4}$ & 1.3\,$\times$\,10$^{-3}$ & 1.9\,$\times$\,10$^{-15}$ & 2.3\,$\times$\,10$^{-15}$ \\
\vspace*{-0.2cm}\\
\hline
\end{tabular}
\end{center}
\end{table*}

The ${CR_\mathrm{max}}$ values were converted into fluxes in physical units ($\mathrm{f_{IC,abs}}$) using the WebPIMMS tool. In each case, a {\tt wabs*power} was assumed, using the N$_\mathrm{H,ISM}$ values quoted in Table\,\ref{nh} and a photon index $\Gamma = 1.5$. Inverse Compton fluxes corrected for the interstellar absorption ($\mathrm{f_{IC}}$) are quoted in the last column of Table\,\ref{ul}.

\section{Discussion}\label{disc}

\subsection{The stars}\label{discstars}

HIP\,16518 does not present any unexpected feature in its X-ray emission deviating from the behaviour of a regular single early B-type star. The flux corrected for interstellar absorption measured in Sect.\,\ref{16518} converts into a luminosity L$_\mathrm{X}$ = 1.2\,$\times$\,10$^{30}$\,erg\,s$^{-1}$. The corresponding L$_\mathrm{X}$/L$_\mathrm{bol}$ ratio is therefore about 10$^{-8}$. This is significantly lower than the $\sim\,10^{-7}$ value expected for single O-type stars. This is in agreement with an expected steeper decline of the X-ray luminosity for the rather weak winds of single B-type star with respect to those of O-type stars \citep{berg1997,owocki2013}.

HIP\,34536 is a rather wide, weakly eccentric binary with a period of about 2100 days. Its orbit was resolved using the Very Large Telescope Interferometer (VLTI) and its first astrometric solution was determined by \citet{vlti2017}. According to this recent ephemeris, our observation corresponds to orbital phase $\phi$\,=\,0.37. Its X-ray emission is the brightest measured in our sample. The X-ray flux corrected for interstellar absorption converts into a luminosity of 2.6\,$\times$\,10$^{32}$\,erg\,s$^{-1}$. Dividing this quantity by the cumulated bolometric luminosities of the O6.5V and O9V companions, one obtains a L$_\mathrm{X}$/L$_\mathrm{bol}$ ratio of about 3\,$\times$\,10$^{-7}$. Such a value does not point to any strong X-ray emission component due to the colliding-winds in the binary. This is not a surprise considering the system is made of mid- and late O-type dwarfs whose winds collide at a rather large distance, thus with a not so high emission measure to feed thermal emission processes. This provides additional support to the idea that wind-wind interaction regions in wide binaries are not necessarily very bright X-ray emitters \citep[see e.g.][]{debecker6604new}.

The late O-star HIP\,77391 is at the limit to be detected in our XMM-Newton observation. Such a weak X-ray emission is most probably explained by a soft emission at a rather low level, significantly absorbed by the interstellar material along the line of sight. On the basis of the count rate and of assumptions on its X-ray emission (see Sect.\,\ref{77391}), the unabsorbed count rate can be converted into a L$_\mathrm{X}$ = 7\,$\times$\,10$^{30}$\,erg\,s$^{-1}$. This value leads to a L$_\mathrm{X}$/L$_\mathrm{bol}$ ratio of about 5\,$\times$\,10$^{-9}$, which is unexpectedly low for an O supergiant. Even though these numbers are very uncertain considering the assumptions made to derive them, the marginal detection in X-rays (only in the pn data set) of HIP\,77391 is surprising. The present information available on this object does not allow to go further in the interpretation of this low emission level.

The early-B star HIP\,78401 (also known as $\delta$\,Sco) is a long period ($\sim$\,10.74\,yr) binary that was intensively studied especially in the optical and near infrared domains \citep[see e.g.][]{delsco1993,delsco2009,delsco2011}. According to the ephemeris published by \citet{delsco2011}, our XMM-Newton observation corresponds to orbital phase $\phi \sim 0.35$. The spectrum is compatible with thermal emission, with an X-ray luminosity corrected for interstellar absorption L$_\mathrm{X}$ = 9.7\,$\times$\,10$^{30}$\,erg\,s$^{-1}$. The L$_\mathrm{X}$/L$_\mathrm{bol}$ ratio of 3.7\,$\times$\,10$^{-8}$ is not so surprising for a B-star.

The X-ray spectrum of the O-type runaway HIP\,97796 agrees well with a thermal nature, with plasma temperatures not larger than about 6\,MK. The flux corrected for interstellar absorption converts into L$_\mathrm{X}$ = 1.74\,$\times$\,10$^{32}$\,erg\,s$^{-1}$, leading to a L$_\mathrm{X}$/L$_\mathrm{bol}$ ratio of 1.0\,$\times$\,10$^{-7}$. The latter value is in full agreement with the X-ray emission from a single O-type star \citep[see e.g.][]{owocki2013}.

\subsection{The bow shocks}\label{discbs}
Even though a few point-like sources may at the limit be considered as potential candidates for the detection of X-ray emission associated with some bow shocks, the nature of these sources could not be clarified with the present data set. For HIP\,34536 (Sect.\,\ref{34536}), one unidentified point source is clearly detected, but the spectral analysis was not conclusive. Close to HIP\,97796 a faint source allowed just for a count rate estimate with no hope for any spectral analysis (Sect.\,\ref{97796}). Finally, in the case of HIP\,16518, a slight X-ray excess is only marginally detected (Sect.\,\ref{16518}). The data available so far could therefore not lead to any unambiguous identification of non-thermal X-rays from bow shocks in our sample. However, the upper limits derived in Sect.\,\ref{ul} deserve to be discussed further in the context of the physics of bow shocks of runaway stars.

The obtained X-ray upper limits can be used to constrain the main physical quantities involved in the non-thermal emission production, using a simple radiation model. We also made use of radio upper limits for the 5 sources we are studying here; in particular we obtained 3-${\sigma}$ radio upper limits for 4 sources using the NRAO all-sky survey at 1.4 GHz \citep[for more information see][]{condon1994}. The lack of archive radio data from  HIP\,77391 did not allow us to compute an upper limit for it; for completeness we use a  value of the same order of that of the other sources. Using the distances listed in Table\,\ref{param} we compute the upper limit radio luminosity at 1.4 GHz. The results are exhibited in the left part of Table\,\ref{upperlimitsresults}. \citet{schulzbsr} derived $\gamma$-ray flux upper limits for 27 stellar bow shocks using data collected by {\it Fermi}, including our 5 targets. The upper limits are calculated in 4 bands covering the energy range from 100\,MeV to 300\,GeV.
  The $\gamma$ upper limits $\mathrm{L_{\gamma,UL}}$, in luminosity units, are also shown in the left part of Table\,\ref{upperlimitsresults}.

\begin{table*}
\caption{Left part: Upper limit luminosities at 1.4 GHz and for 4 $\gamma$-ray bands defined in GeV. See the text for further details. Right part: Free parameters and optimal values. \label{upperlimitsresults}}
\begin{center}
\begin{tabular}{l c c c c c c c c c c c c}
\hline
Target & \multicolumn{5}{c}{Upper limits} & & \multicolumn{6}{c}{Optimal values}\\
\cline{2-6}\cline{8-13}
\vspace{0.1cm}
 & $\mathrm{L_{RD,UL}}$ & \multicolumn{4}{c}{$\mathrm{L_{\gamma,UL}}$ [(10$^{35}$\,erg\,s$^{-1}$)]} & & $B$  & $\chi_{\rm rel}$ & $\alpha$ & $E_{\rm max}$ & $\chi_{\rm IR}$& $\mathcal{D}$ \\
 & (erg\,s$^{-1}$) & 0.1-0.74 & 0.74-5.5 & 5.5-41 & 41-300 & & (G)  &  &  & ($m_ec^2$)    &  &  \\
\hline
HIP\,16518 & 1.4\,$\times$\,10$^{27}$ & 0.74 & 0.16\ & 0.16 & 1.0 & & $10^{-4}$ & 1.0 & 1.8 & 5.3$\times 10^{3}$ & 1.0 & 5.2$\times 10^{-1}$ \\
HIP\,34536 & 5.6\,$\times$\,10$^{27}$ & 1.6 & 2.0 & 1.4 & 6.2 & & 9$\times 10^{-6}$ & 0.3 & 2.1 & 5.3$\times 10^{5}$ & 0.4 & 7.0$\times 10^{-3}$ \\
HIP\,77391 & 2.1\,$\times$\,10$^{27}$ & 1.7 & 5.9 & 6.8 & 1.3 & & 2.4$\times 10^{-5}$ & 0.1 & 2.1 & 1.1$\times 10^{6}$ & 0.1 & 8.1$\times 10^{-3}$ \\
HIP\,78401 & 1.7\,$\times$\,10$^{26}$ & 0.055 & 0.014 & 0.032 & 0.1 & & 2.1$\times 10^{-5}$ & 0.07 & 2.5 & 8.2$\times 10^{5}$ & 0.06 & 1.1$\times 10^{-2}$ \\
HIP\,97796 & 1.2\,$\times$\,10$^{28}$ & 7.9 & 3.3 & 3.2 & 9.1 & & 1.4$\times 10^{-5}$ & 0.1 & 2.1 & 3.7$\times 10^{5}$ & 0.2 & 8.1$\times 10^{-3}$ \\
\hline
\end{tabular}
\end{center}
\end{table*}

The approach adopted here is not strictly speaking a regular fit procedure, as we are dealing only with upper limits and not actual measurements. In the absence of detection of non-thermal radio and X-ray emission from our targets, one has to follow a two-step approach. First of all, we use a simple radiation model to determine for which set of physical parameters (within allowed ranges) the predicted non-thermal luminosities are as close as possible to the upper limits. Second, the derived set of parameters must be critically discussed to identify the likely reasons for the non-detection of our targets. The radiation model and the results of our approach are described below.

\subsubsection{The radiation model}\label{rad}
We assume that a population of relativistic electrons exists in each source, with energy distribution $N(E)$ in the interval $E_{\rm min} \leq E \leq E_{\rm max}$, where $E$  is the particle energy. The electrons distribution is approximated as:

\begin{equation}
N(E) \sim Q_0 E^{-\alpha} t_{\rm loss},
\end{equation}
\noindent where $Q_0$ is a normalization constant and $t_{\rm loss}$ is the time scale of the relevant losses: radiative and non-radiative losses. The most important radiative losses are caused by infrared photons inverse Compton scattered off by the relativistic electrons and synchrotron radiation. The non-radiative losses are caused by advection, that draws the particles away from the acceleration region where they would be able to radiate \citep{ntbowshock}. Hence $t_{\rm loss} = {\rm min}(t_{\rm IC}, t_{\rm synchr}, t_{\rm adv})$. The value of $Q_0$ is directly related to the available power in the system ${P_\mathrm{kin}^{BS}}$ to accelerate particles (i.e. the kinetic energy budget in the system):
  
\begin{equation}
Q_0  = \frac{1}{\Omega}{L_{\rm rel}}\, \left( {\int^{E_{\rm max}}_{E_{\rm min}} E^{-\alpha+1}{\rm d}E} \right)^{-1} ,
\end{equation}
\noindent where $L_{\rm rel} = \chi_{\rm rel}\mathrm{P_{kin}^\mathrm{BS}}$, and $\chi_{\rm rel}$ is the fraction of the available power injected into relativistic particles (i.e. the shock efficiency) and $\Omega$ is the emission volume. The wind kinetic power is:
\begin{equation}\label{pkin}
{P_\mathrm{kin} = \frac{1}{2}\,{\dot M}\,V_\infty^2}.
\end{equation}
Only a fraction ${f_\mathrm{BS}}$ of this kinetic power is directed towards the bow shock. This fraction can be estimated as the ratio of the surface of the bow shock (modelled as a spherical cap with thickness $w$) to the surface of a sphere of radius $R$ (see Table~\ref{param}), and assuming further the sphere and the spherical cap share the same centre:

\begin{equation}\label{fbs}
{f_\mathrm{BS} = \frac{2\pi\,R\,(w/2)}{4\pi\,R^2} = \frac{w}{4\,R}}.
\end{equation}

\noindent Hence, $L_{\rm rel} = \chi_{\rm rel} {P_\mathrm{kin}^\mathrm{BS}} = \chi_{\rm rel} {f_\mathrm{BS}} {P_\mathrm{kin}}$; the values of ${P_\mathrm{kin}^\mathrm{BS}}$ for each source are given in Table~\ref{ntbud}. Further discussion on the energy budget of the system is developed in Sect.\,\ref{enbud}. 

As shown in previous studies \citep[e.g.][]{ntbowshock}, the non-thermal spectrum should be dominated by  synchrotron emission, produced by electron interaction with some local magnetic field $B$, and IC radiation, resulting from collisions with IR photons. 

The IR photons are emitted by swept-up dust heated by the stellar photons \citep{vanburen1988}. The IR luminosity emitted by the dust in the bow shock is a fraction of the bolometric luminosity, with  typical  fractions of $\chi_{\rm IR} = 10^{-2}$ \citep[e.g.,][]{vanburen1988}. The density of target photons $n_{\rm targ}$ for the IC process is calculated considering a black body at $T = 40$~K (i.e. a characteristic value for stellar bow shocks) with a normalization factor to a fraction of the bolometric luminosity (some authors call this a grey body). Let $L_{\rm targ}$ be the luminosity associated with the photon  target density, then the correction factor is such that $L_{\rm targ} = \chi_{\rm IR}$L$_\mathrm{bol}$. For each source we use the values of  $L_\mathrm{bol}$ listed in Table\,\ref{param}.

In order to compute the synchrotron and IC emissivities per unit energy ($q_{\epsilon}^{\rm synchr}$ and $q_{\epsilon}^{\rm IC}$, respectively)  we use the general formulae given by \citet{BG1970}.

The fundamental parameters in this simple model for the non-thermal emission are: the magnetic field $B$, the power in relativistic electrons, in this case as a fraction ($\chi_{\rm rel}$) of the available power in the bow shock, the maximum energy of the electron distribution $E_{\rm max}$, the injection power-law index $\alpha$ and the number of target IR photons as a function of the bolometric luminosity $\chi_{\rm IR}$. All these values can be reasonably constrained using physical arguments. We aim at finding the most suitable parameters (within given intervals) that best accommodate with the derived upper limits both in the radio and X-ray domains\footnote{We do not use the $\gamma$ luminosity upper limits because they are too poorly constrained when compared with the X-ray upper limits.}. In the next subsection we describe the optimization method we implemented.

\subsubsection{Optimization method and application}\label{fit}
We use a cross-entropy method for variance minimization. This is  an iterative method that uses a random sampling of the free parameters in a given interval. For details on the method the reader is referred to  \citet{deBoer2005}. The result is a set of values for the parameters that best fit the variance minimization condition\footnote{The result might not be unique, hence this method cannot distinguish local minima from a global one, however given the uncertainties involved in the modelling this is enough for our purposes.}.

In our case, as we are interested in comparing the order of magnitude of the theoretical luminosities calculated with our model with the maximum luminosities allowed by the upper limits, we minimize the following function:
\begin{equation}\label{distance}
   \mathcal{D} = \sqrt{(L_{\rm synchr}^{1.4\,{\rm GHz}}/\mathrm{L_{RD,UL}} - 1)^2 + ({L}_{\rm IC}/\mathrm{L_{IC,UL}} - 1)^2},
\end{equation}
\noindent where $L_{\rm synchr}^{1.4\,{\rm GHz}}$ is the theoretical synchrotron radiation at frequency $1.4\,{\rm GHz}$ ($\sim$ $5.8\times 10^{-6}$\,eV) and ${L}_{\rm IC}$ is the theoretical IC luminosity in the range 0.3 and 10 keV. For computing ${L}_{\rm IC}$ we integrate the emissivity in the corresponding energy band, i.e.
\begin{equation}\label{lumi}
{L}_{\rm IC} = \int_{0.3\,{\rm keV}}^{10\,{\rm keV}}\epsilon \, q_{\epsilon}^{\rm IC}\,{\rm d}\epsilon;
\end{equation}
while $L_{\rm synchr}^{1.4\,{\rm GHz}}$ is simply  $E_{\rm synchr}^{2}\times q_{\epsilon}^{\rm synchr}(E_{\rm synchr})$, with $E_{\rm synchr} = 5.8\times 10^{-6}$\,eV.

The free parameters are $B$, $\chi_{\rm rel}$, $\alpha$, $E_{\rm max}$, and $\chi_{\rm IR}$. The corresponding intervals are listed in Table\,\ref{intervals}. We adopt a magnetic field of $\sim$ 1 ${\mu}$\,G (of the order of the ISM magnetic field) as a reference value\footnote{Making stronger assumptions on the magnetic field is out of the scope of our simple radiation model.} and from that we consider values  2 orders of magnitude greater and lower. The minimum magnetic field in the interval can be used as a diagnostic tool. If the best fitted value is near this lower limit it means that efficient shock acceleration is not possible and therefore non-thermal emission would not be produced. The maximum value in the interval can be achieved in principle by non-linear effects during shock acceleration when, in the presence of turbulence, the ambient magnetic field can be amplified, reaching high values \citep[e.g.,][]{delvalle16}. The linear theory of DSA leads to a power-law injection index $\alpha$ of $\sim$ 2, although non-linear and other effects might result in deviations from this value \citep[e.g.,][]{longairbook}. For the maximum value
  in the $E_{\rm max}$ interval  we adopt  $10^{7}\,m_ec^2$ based on the results of \citet{ntbowshock}. All parameters but $\alpha$ vary by orders of magnitude, so logarithm intervals are used throughout our optimization approach.

\begin{table}
\caption{Free parameters of the radiative model and the corresponding intervals. \label{intervals}}
\begin{center}
\begin{tabular}{l c}
\hline
Parameter & interval\\
\hline
$B$ [G] & ($10^{-8}$, $10^{-4}$)\\
$\chi_{\rm rel}$ & ($10^{-2}$, 1)\\
$\alpha$ & (1.8, 4)\\
$E_{\rm max}$ [$m_ec^2$] & (10, $10^{7}$)\\
$\chi_{\rm IR}$ & ($10^{-2}$, 1)\\
\hline
\end{tabular}
\end{center}
\end{table}

\subsubsection{Results}\label{res}

The optimal set of parameter values and the value of $\mathcal{D}$ (see Eq.\,\ref{distance}) for each source are shown in the right part of Table\,\ref{upperlimitsresults}. In all cases but HIP\,16518, the theoretical models accommodate well within the observational constraints, with $\mathcal{D} \leq 1.0\times 10^{-2}$, meaning that the model luminosities and the upper limits are of the same order.

For all sources but HIP\,16518 (discussed separately below) the parameter values appear quite reasonable. The spectral indices are $\sim$ 2, as expected from a DSA mechanism. The shock efficiencies, $\chi_{\rm rel}$, are of the order expected in this process. For these four sources the maximum electron energies are between 30 and 100 GeV, allowing for the IC emission to reach the $\gamma$-ray energy band. The magnetic field values required to reach the radio upper limits are of the order of $\sim$ $10^{-5}$\,G, the same order found for BD$+$43$^\circ$3654 by \citet{benagliarunaway}. Only for source HIP\,78401 the value of $\chi_{\rm IR}$ is lower than 0.1. The IR luminosity required for reaching the IC upper limit for that source is at least one order of magnitude lower than for other sources. However, all values are reasonable considering that we assume that all the bow shocks emit at the same temperature. Also notice that the bow shock dust emission depends on many factors -- such as the dust model, density distribution, stellar photons transport, etc. -- that are ignored here.

\begin{figure}
\begin{center}
\includegraphics[scale=.35,trim = 3.cm 1.5cm 2.7cm 1.5cm, angle=270]{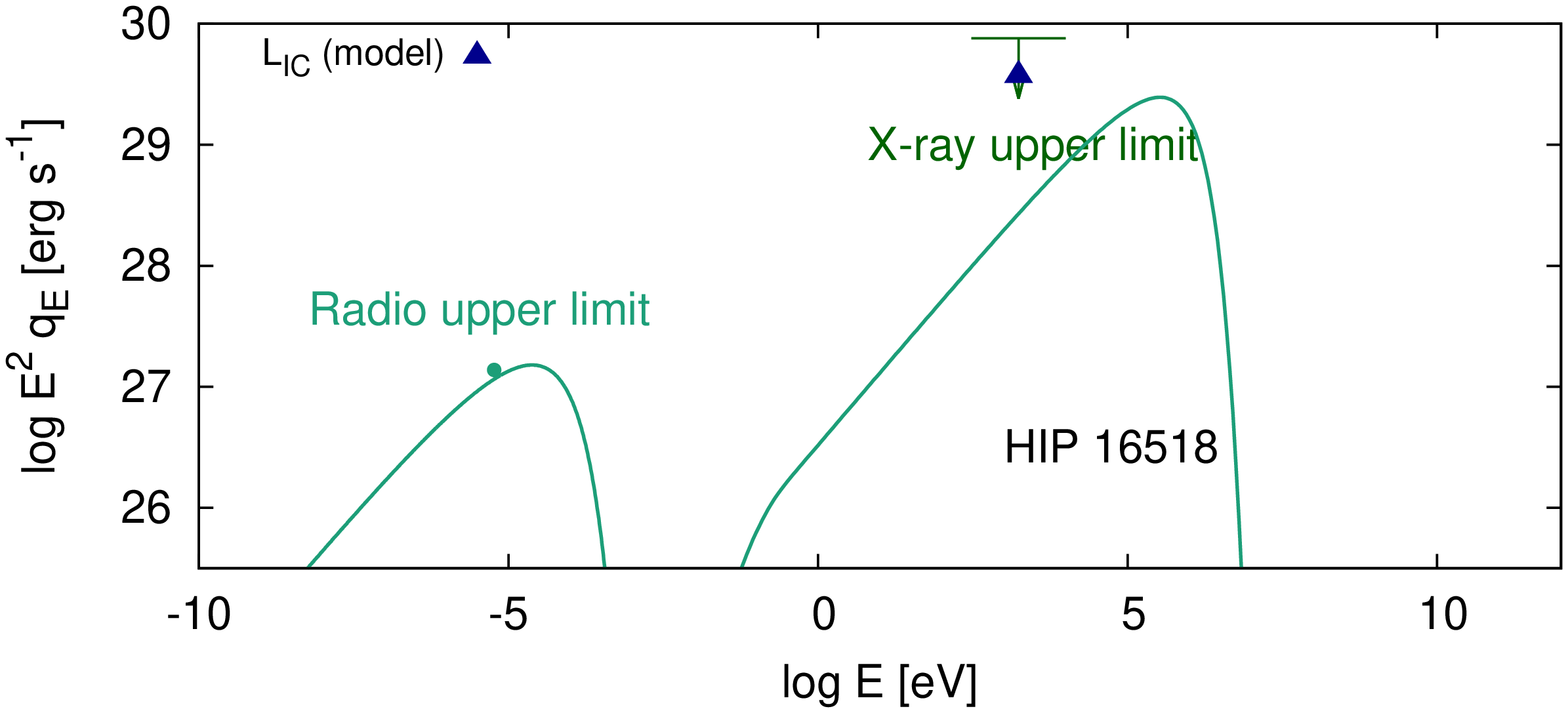}\\
\includegraphics[scale=.35,trim = 3.cm 1.5cm 2.7cm 1.5cm, angle=270]{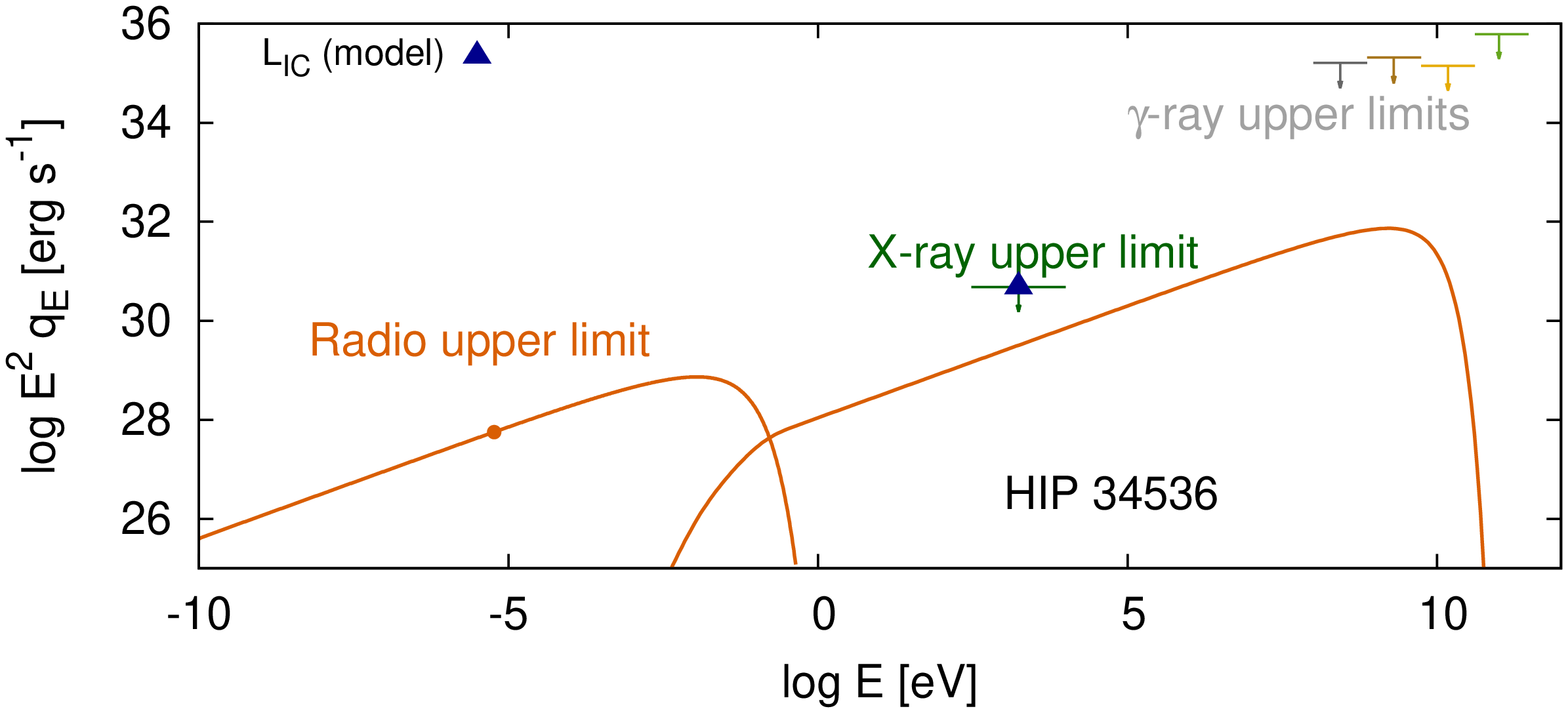}\\
\includegraphics[scale=.35,trim = 3.cm 1.5cm 2.7cm 1.5cm, angle=270]{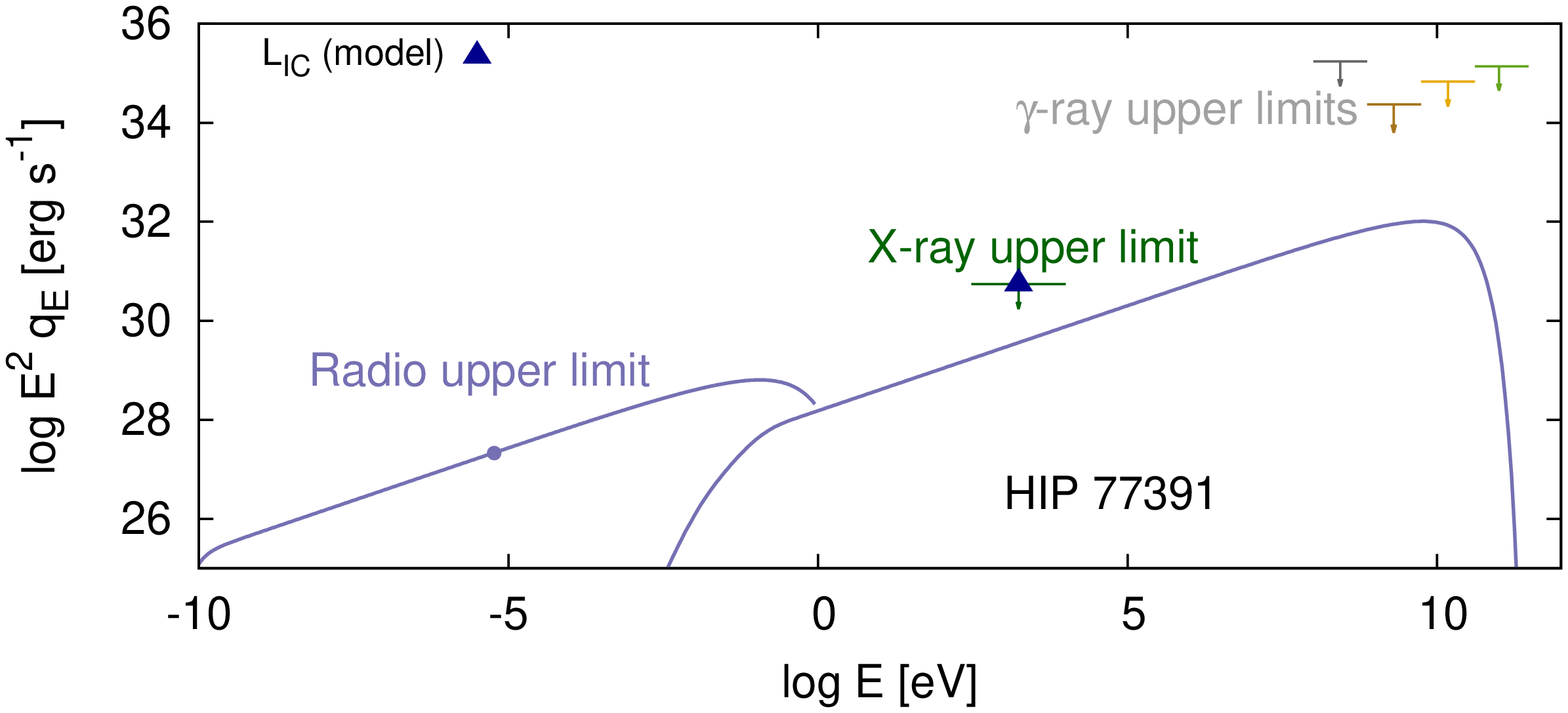}\\
\includegraphics[scale=.35,trim = 3.cm 1.5cm 2.7cm 1.5cm, angle=270]{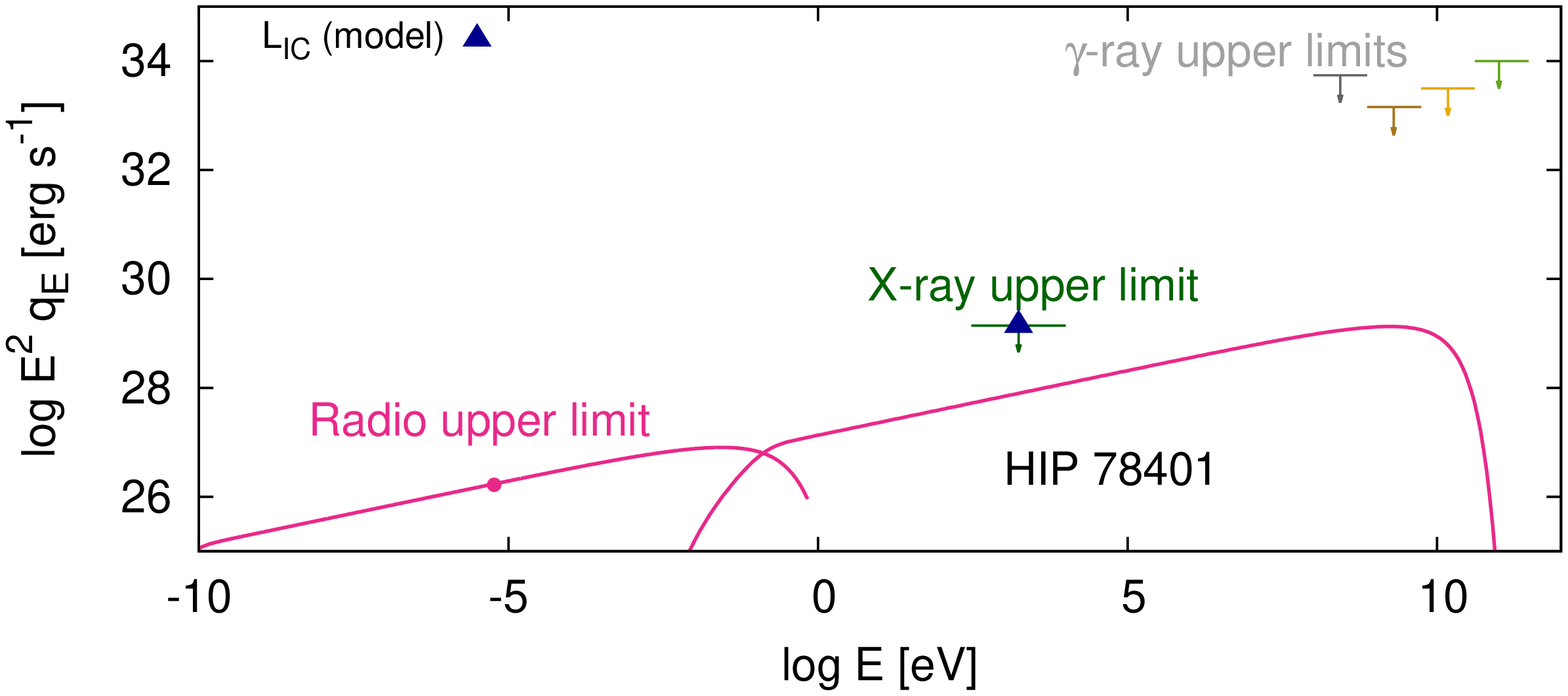}\\
\includegraphics[scale=.35,trim = 3.cm 1.5cm 2.7cm 1.5cm, angle=270]{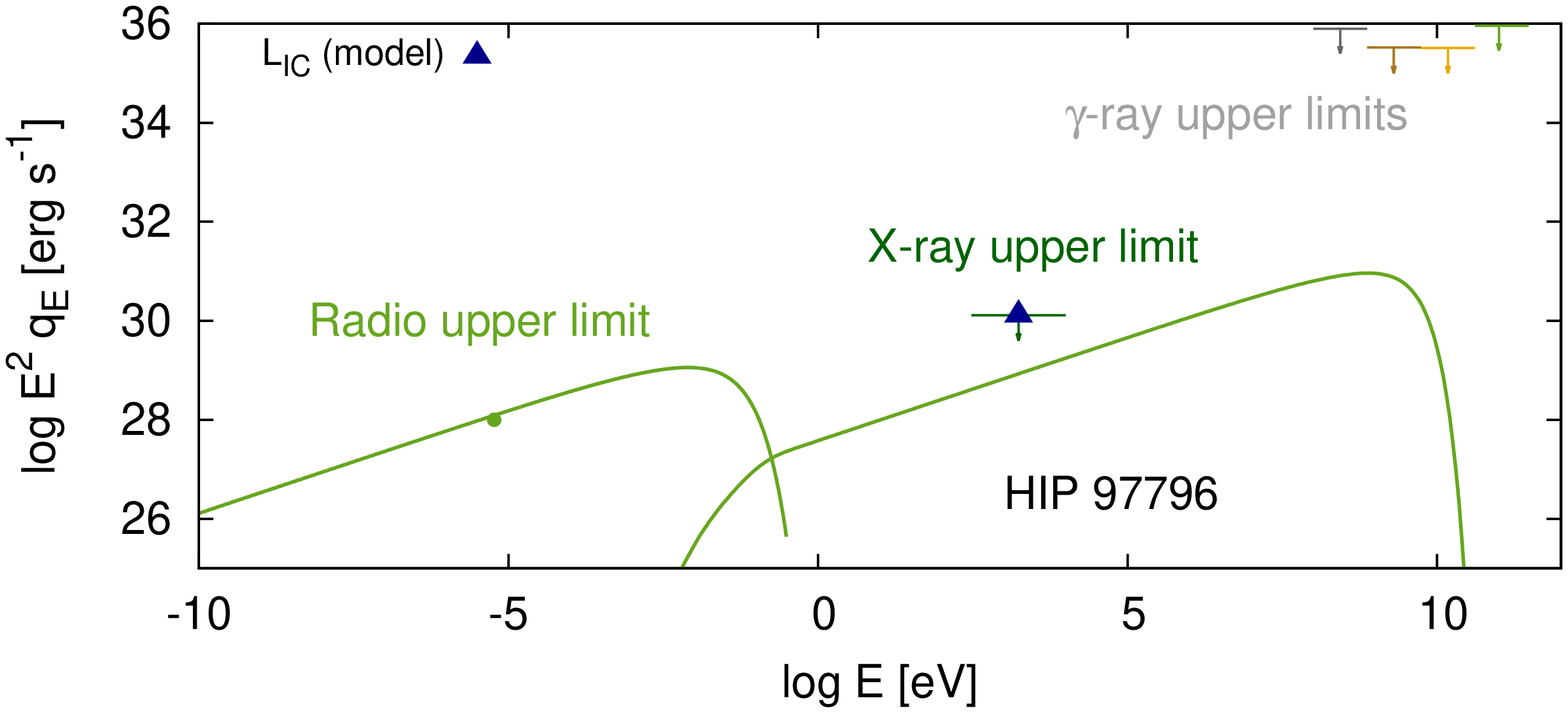}

\caption{Synchrotron and IC luminosity per unit energy for the 5 sources. These spectral energy distributions correspond to the best-set of parameter values allowed by the upper limits for each target (shown in the right part of Table\,\ref{upperlimitsresults}). The  blue triangles correspond to the integrated luminosities $L_{\rm IC}$. The colored dots indicate the radio upper limits. The arrows indicate the upper limits in the X-ray band (from 0.3 to 10 keV) and in 4 $\gamma$-ray bands (from 100 MeV to 300 GeV).} \label{seds}
\end{center}
\end{figure}

The model gives an  X-ray integrated luminosity for HIP\,16518 ${L}_{\rm IC} = 3.7\times 10^{29}$\,erg\,s$^{-1}$, i.e. 50\% below the upper limit. The synchrotron luminosity at 1.4\,GeV is however of the same order of $\mathrm{L_{RD,UL}}$ (see the dot and the blue triangle in the top plot of Fig.\,\ref{seds}). For reaching these luminosities the magnetic field, $\chi_{\rm rel}$, and $\chi_{\rm IR}$ adopt the maximum values of their allowed intervals. This result is not surprising given that this is the less energetic target. For this reason it is likely that this source does not produce significant non-thermal radiation. 
 In addition, this source gives a rather hard injection index, $< 2$, not expected in the shocks of stellar winds. The maximum energy lays two orders of magnitude below the maximum energies of the other cases. Altogether, we conclude that this source is an inefficient non-thermal emitter.

In Figure\,\ref{seds} we show the computed non-thermal luminosity per unit energy (i.e., non-thermal emissivity per unit energy multiplied by E$^2$) for the five sources. The radio upper limits are marked with a dot. We also indicate with arrows the X-ray and $\gamma$-ray (GeV) upper limits, even though these are quantities integrated in energy. The theoretical X-ray integrated luminosities ${L}_{\rm IC}$ are also shown as blue triangles. Notice that the luminosity upper limits are the power integrated over a given energy band (see Eq.\,(\ref{lumi})), while the curves in Fig.\,\ref{seds} represent the power as a function of energy. Even when both quantities have the same units the curves are not supposed to match the luminosity upper limits, because they represent different quantities (in fact the integrated luminosity is expected to be higher than the power as a function of energy). The case of the synchrotron radiation is special in the sense that the luminosity upper limit is given at a single energy and not over a band. It coincides with $E^{2}\times q_{\epsilon}$ at $E_{\rm synchr}$. In the case of source HIP\,16518, the $\gamma$-ray upper limits are not shown because the emission does not reach such high energies. The higher energy $\gamma$-ray (TeV) upper limits derived by \citet{abdo2013} are not shown. As expected the spectral energy distributions for the cases HIP\,34536 to HIP\,97796 are similar. The IC emission extends up to energies of the order of a few GeVs, with the maximum values reaching around $\sim$ 0.1\, GeV. In all cases the $\gamma$-ray upper limits are many orders of magnitude over the obtained luminosities.

According to our model, if we assume that electrons are accelerated with injection index $\sim$ 2 and for fixed maximum electron energies of several tens of GeV\footnote{This assumption is reasonable given that we obtained similar and very acceptable values for these parameters from all sources.} (i.e., fixing $E_{\rm max}$ and $\alpha$), the values we obtained for  $B$, $\chi_{\rm rel}$ and  $\chi_{\rm IR}$ (listed in the right part of Table\,\ref{upperlimitsresults}) can be considered as upper limits. Higher values of these parameters would indeed produce emission greater than the observational upper levels at X-rays and radio wavelengths. Also, under these conditions, the maximum $\gamma-$ray radiation occurs in the soft $\gamma$-ray domain, around $1-10$\,GeV, with a maximum achievable luminosity of the order of $10^{32}$\,erg\,s$^{-1}$ (for HIP\,77391).

In principle, all targets but HIP\,16518 might produce non-thermal emission at the level of the radio and X-ray upper limits. However, we emphasize that we are considering a simple physical model. Many factors might cause a lower non-thermal luminosity. For instance, particles might not be efficiently  accelerated in the forward or reverse shock. This can happen under several circumstances: when the shocks are too slow (this also affects $\chi_{\rm rel}$), also if the magnetic field is too weak, when the material in the pre-shock region is not fully ionized or too dense, in the case of stronger advection losses that might be catastrophic for particle acceleration, etc. In addition, the relativistic electrons might diffuse too slowly to the region where the density of target photons is higher. 

Another possibility is that the energy density of IR photons in the bow shock is not high enough to produce a detectable IC signal. Note that the values we obtained for $\chi_{\rm IR}$, though theoretically possible, are above the typical observed ones. Concerning the magnetic field, even though the radio detection of BD$+$43$^\circ$3654 implies a high magnetic field in the source, it is not clear whether such values are expected in the surroundings of the reverse shock of most of sources (the most efficient site for DSA). A deeper discussion of these and other caveats is out of the scope of this work and will be presented elsewhere.

\subsection{Energy budget}\label{enbud}

It can be instructive to set the physical quantities used in this paper into the general framework of a discussion of the energy budget. Figure\,\ref{EnBud} is a chart that allows to follow the energy flux from its initial pool (the stellar radiation field) up to the non-thermal processes. A short description of the various energy transfers represented by the numbered arrows is given below:

\begin{figure*}
\begin{center}
\includegraphics[width=14cm]{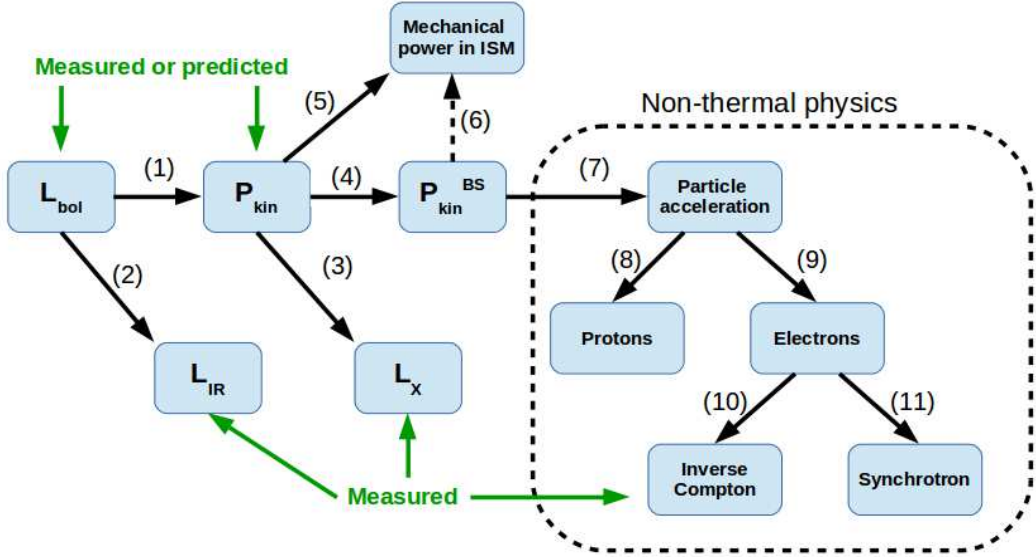}
\caption{Schematic view of the energy budget of bow shock runaways with emphasis on the non-thermal physics. \label{EnBud}}
\end{center}
\end{figure*}

\begin{enumerate}
\item[1.] The stellar wind is driven by the radiation pressure due to the intense radiation field from the stellar photosphere. A fraction of the bolometric luminosity is thus converted into mechanical power in the stellar wind. The starting point is thus the bolometric luminosity, which can for instance be predicted on the basis of the stellar classification of the object. The mechanical power injected in the stellar wind is represented by its kinetic power ${P_\mathrm{kin}}$, see Eq.\,(\ref{pkin}).
%
%
\noindent It is easily estimated on the basis of measurements and/or predictions on the mass loss rate and the terminal velocity. The stagnation radius of bow shocks is much larger than the wind acceleration length-scale (at most a few stellar radii). Stellar winds have therefore reached their asymptotic velocities at the bow shock location. Typically, a fraction of 10$^{-4}$ to 10$^{-6}$ of L$_\mathrm{bol}$ is converted into ${P_\mathrm{kin}}$, depending on the stellar category.
\item[2.] The heating of material, and especially dust particles, by the stellar radiation field leads to the production of thermal infrared emission. This radiation is measured and allows for the detection of the bow shocks \citep{ebossI,ebossII}.
\item[3.] The thermal X-ray emission from stellar winds constitutes only a minor fraction of the wind kinetic power. Typically, values of the order of a 0.1--1$\%$ are measured.
\item[4.] A fraction of the kinetic power is directed toward the bow shock revealed by infrared observations. The fraction ${f_\mathrm{BS}}$ going in the right direction can be estimated on the basis of geometrical considerations as done previously  (see Eq.\, \ref{fbs}).  The kinetic power directed toward the bow shock is
\begin{equation}\label{pkinbs}
{P_\mathrm{kin}^\mathrm{BS} = f_\mathrm{BS}\,P_\mathrm{kin}}
\end{equation}
\item[5.] The fraction of the kinetic power not directed toward the bow shock is directly injected into the interstellar medium.
\item[6.] A significant fraction of $\mathrm{P_{kin}^{BS}}$ is used in the advection of material along the shocks and goes to the interstellar medium.
\item[7.] A fraction of the kinetic power injected in the bow shock drives DSA. This is the basic injection of energy into non-thermal particles.
\item[8.] Part of the energy invested in the DSA will go into relativistic protons and other nuclei. Most of these particles will escape and contribute to the low energy part of the Galactic cosmic ray population. 
\item[9.] The remaining energy goes essentially to electrons.
\item[10.] Relativistic electrons are expected to be mainly cooled down by inverse Compton scattering of infrared photons produced by warm dust, as illustrated in spectral energy distributions and cooling-time plots presented in \citet{ntbowshock}, \citet{DRD2013} and \citet{toala2016}.
\item[11.] Finally, a fraction of the energy injected into relativistic electrons is converted into synchrotron emission, especially in the radio domain.
\end{enumerate}

Some values of the physical quantities participating in the energy budget are given in Table\,\ref{ntbud}. Some of them were determined on the basis of Eq.\,\ref{pkin}, \ref{fbs} and \ref{pkinbs}, using parameters given in Table\,\ref{param}. The upper limit on the IC luminosity (between 0.3 and 10.0\,keV) is determined on the basis of flux values given in the last column of Table\,\ref{ul}. The upper limit on the efficiency ($\mathrm{\eta_{IC}}$) of the conversion of kinetic power (${P_\mathrm{kin}^\mathrm{BS}}$) into IC radiation ($\mathrm{L_{IC,UL}}$) is also given. The latter quantity can be viewed as an indicator of the maximum efficiency of bow shock runaways to produce inverse Compton radiation in X-rays, according to our upper limit determinations.

\begin{figure}
\begin{center}
\includegraphics[width=8cm]{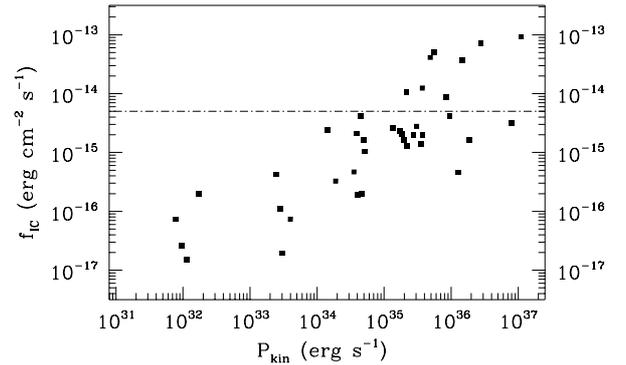}
\caption{Plot of the anticipated upper limit on the inverse Compton fluxes in the soft X-ray band ($\mathrm{f_{IC}}$) as a function of the stellar wind kinetic power (${P_\mathrm{kin}}$) for 39 members of the E-BOSS catalogue, assuming an inverse Compton efficiency ($\mathrm{\eta_{IC}}$) equal to 10$^{-5}$. The horizontal line represents the typical level of the faintest sources detected with XMM-Newton. \label{bsric}}
\end{center}
\end{figure}

These considerations are useful to discuss our expectations for the detection of inverse Compton X-rays from bow shocks extrapolated to a more extended sample. We therefore considered all the members of the E-BOSS catalogue \citep{ebossI,ebossII}. We first determined ${P_\mathrm{kin}}$ for all objects with known ${\dot M}$ and $V_\infty$. Every time it was possible, we estimated ${f_\mathrm{BS}}$ (see Eq.\,\ref{fbs}) to determine ${P_\mathrm{kin}^\mathrm{BS}}$. On the basis of the results obtained for the sample studied in this paper (see Table\,\ref{ntbud}), we assumed a value for $\mathrm{\eta_{IC}} = 10^{-5}$ to estimate pessimistic values for the upper limit on $\mathrm{L_{IC}}$. Finally, provided the distance was known, we derived corresponding values for the upper limits on the X-ray flux $\mathrm{f_{IC}}$. We note that, in the context of the absorbed power law model used in Sect.\,\ref{upperlimits}, absorbed and unabsorbed fluxes are very similar, and values of $\mathrm{f_{IC}}$ are good indicators of measured (i.e. absorbed) fluxes as well. Figure\,\ref{bsric} plots the $\mathrm{f_{IC}}$ values as a function of ${P_\mathrm{kin}}$. A general trend suggests, as expected, that runaways with higher kinetic power leads to higher level values of $\mathrm{f_{IC}}$. One sees that for most potential targets, upper limits on fluxes are of the order of magnitude -- or even below -- the actual order of magnitude of the faintest X-ray sources detected with XMM-Newton, i.e. $\sim 5\,\times\,10^{-15}$\,erg\,cm$^{-2}$\,s$^{-1}$ \citep[see e.g.][]{XMMSS3}. As one is talking about upper limits, points above the horizontal line in Fig.\,\ref{bsric} can not be interpreted in terms of likely detection. In addition, one has to keep in mind that any putative detection is still not enough to clarify the nature (thermal vs non-thermal) of the emission. 

As a result, we cannot anticipate a priori that observations of runaways with bow shocks could lead to clear identifications of non-thermal X-ray emission with present X-ray observatories. At least one order of magnitude improvement in the sensitivity is required, postponing the prospects in this field to the Advance Telescope for High Energy Astrophysics (Athena) observations, expected for the next decade \citep{athena2015}.  

\begin{table}
\caption{Physical quantities playing a role in the energy budget illustrated by Fig.\,\ref{EnBud}. \label{ntbud}}
\begin{center}
\begin{tabular}{l c c c c}
\hline
Target & ${P_\mathrm{kin}}$ & ${P_\mathrm{kin}^{BS}}$ & $\mathrm{L_{IC,UL}}$ & $\mathrm{\eta_{IC}}$ \\
 & (erg\,s$^{-1}$) & (erg\,s$^{-1}$)& (erg\,s$^{-1}$) & \\
\hline
\vspace*{-0.2cm}\\
HIP\,16518 & 4.7\,$\times$\,10$^{32}$ & 1.7\,$\times$\,10$^{32}$ & 7.6\,$\times$\,10$^{29}$ & 4\,$\times$\,10$^{-3}$ \\
HIP\,34536 & 3.6\,$\times$\,10$^{35}$ & 6.8\,$\times$\,10$^{34}$ & 4.8\,$\times$\,10$^{30}$ & 7\,$\times$\,10$^{-5}$ \\
HIP\,77391 & 3.1\,$\times$\,10$^{35}$ & 7.8\,$\times$\,10$^{34}$ & 5.5\,$\times$\,10$^{30}$ & 7\,$\times$\,10$^{-5}$ \\
HIP\,78401 & 5.3\,$\times$\,10$^{34}$ & 4.2\,$\times$\,10$^{33}$ & 1.4\,$\times$\,10$^{29}$ & 3\,$\times$\,10$^{-5}$ \\
HIP\,97796 & 6.2\,$\times$\,10$^{35}$ & 6.2\,$\times$\,10$^{34}$ & 1.3\,$\times$\,10$^{30}$ & 2\,$\times$\,10$^{-5}$ \\
\vspace*{-0.2cm}\\
\hline
\end{tabular}
\end{center}
\end{table}

\section{Summary and conclusions}\label{concl}

We report on the analysis of XMM-Newton observations of five massive runaways with bow shocks. The main objective was to investigate the possibility that they could reveal non-thermal X-ray emission, in line with recent theoretical predictions. The careful inspection of all data sets did not reveal the presence of any X-ray emission spatially coincident with the bow shocks. On the other hand, the X-ray emission for the stars is discussed and interpreted in terms of the standard thermal emission from stellar winds.

On the basis of measurements of the background level in regions devoid of point sources, we determined conservative count rate upper limits on the putative X-ray emission at the position of the bow shocks. Assuming a power-law model adequate for inverse Compton scattering of infrared photons produced by heated dust, we converted these upper limits into physical flux units. In parallel, we used archive radio data to derive upper limits on the synchrotron radio emission from the bow shocks.\\

We used a simple radiation model for computing the theoretical synchrotron and inverse Compton emission from the bow shocks to be confronted to X-ray and radio upper limits. To do so, we first used the upper limits as actual values. In this way we derived constraints on model parameters involved in the non-thermal radiation production. Our results suggest that one of the targets is not able to produce significant non-thermal radiation. The set of parameters (magnetic field, amount of incident IR photons, etc) obtained for the other 4 sources lay within physically reasonable values, suggesting a priori they could in principle emit at a level close to the observational upper limits. Our models allow to convert the observational upper limits into upper limits on some relevant physical parameters. 

In particular, we estimated the efficiency of the shocks and the injection power-law index. The obtained values are in agreement with the expected values for DSA in strong shocks. The local magnetic field allowed by the upper limits is of the order of $10$\,$\mu$G, higher than the ISM magnetic field. Assuming that electrons can be effectively accelerated in the sources through DSA, the reason why the non-thermal emission lays under the detection levels might be because the magnetic field is lower than this value and/or because the number of IR photons in the source is not high enough. In addition, one can not reject the idea that relativistic electrons may diffuse too slowly to the region where the density of target photons reaches its highest value, therefore preventing the non-thermal emission process to operate significantly.

Finally, on the basis of energy budget considerations we discussed these non-detections into a wider context, and extrapolated the behavior of the population of bow shock runaways in the E-BOSS catalogue. This approach -- with the support of our modelling --  allows to anticipate that the detection of inverse Compton scattering emission from the bow shock of massive runaways with current X-ray observatories is very unlikely. The advent of next generation X-ray space observatories such as Athena constitutes a strong requirement to pursue the investigation of the high energy non-thermal emission from these objects.

\section*{Acknowledgements}
The authors want to warmly thank people working at the XMM-SOC for the scheduling of the XMM-Newton observations. M. De Becker acknowledges the financial support from ULg through a 'Cr\'edit classique (DARA)'. The SIMBAD database has been consulted for the bibliography. G.~E. Romero acknowledges support from the Argentine Agency CONICET (PIP 2014-00338) and the Spanish Ministerio de Econom\'\i a y Competitividad (MINECO/FEDER, UE) under grants AYA2013-47447-C3-1-P and AYA2016-76012-C3-1-P. M.~V. del Valle acknowledges support from the Alexander von Humboldt Foundation. Finally, the authors want to thank the anonymous referee for a careful reading and for constructive comments that helped to improve the paper.



\bibliographystyle{mnras}

\begin{thebibliography}{99}

\bibitem[\protect\citeauthoryear{{Abdo}, {Ajello}, {Allafort} \& {et~al.}}{{Abdo} et~al.}{2013}]{abdo2013}
{Abdo} A.~A., {Ajello} M., {Allafort} A., {et~al.} 2013, \apjs, 208, 17

\bibitem[\protect\citeauthoryear{{Arnaud}}{{Arnaud}}{1996}]{xspec1996}
{Arnaud} K.~A.,  1996, in {Jacoby} G.~H.,  {Barnes} J.,  eds, Astronomical Data
  Analysis Software and Systems V Vol.~101 of Astronomical Society of the
  Pacific Conference Series, {XSPEC: The First Ten Years}.
p.~17

\bibitem[\protect\citeauthoryear{{Barcons}, {Nandra}, {Barret}, {den Herder},
  {Fabian}, {Piro}, {Watson} \& {the Athena Team}}{{Barcons}
  et~al.}{2015}]{athena2015}
{Barcons} X.,  {Nandra} K.,  {Barret} D.,  {den Herder} J.-W.,  {Fabian} A.~C.,
   {Piro} L.,  {Watson} M.~G.,    {the Athena Team} 2015, Journal of Physics
  Conference Series, 610, 012008

\bibitem[\protect\citeauthoryear{{Bedding}}{{Bedding}}{1993}]{delsco1993}
{Bedding} T.~R., 1993, \aj, 106, 768

\bibitem[\protect\citeauthoryear{{Bell}}{{Bell}}{1978}]{bell1978}
{Bell} A.~R., 1978, \mnras, 182, 147

\bibitem[\protect\citeauthoryear{{Benaglia}, {Romero}, {Mart{\'{\i}}}, {Peri}
  \& {Araudo}}{{Benaglia} et~al.}{2010}]{benagliarunaway}
{Benaglia} P.,  {Romero} G.~E.,  {Mart{\'{\i}}} J.,  {Peri} C.~S.,    {Araudo}
  A.~T.,  2010, \aap, 517, L10
  
\bibitem[\protect\citeauthoryear{{Berghoefer}, {Schmitt}, {Danner} \&
  {Cassinelli}}{{Berghoefer} et~al.}{1997}]{berg1997}
{Berghoefer} T.~W.,  {Schmitt} J.~H.~M.~M.,  {Danner} R., {Cassinelli} J.~P.,  1997,
  \aap, 322, 167  

\bibitem[\protect\citeauthoryear{{Blumenthal} \& {Gould}}{{Blumenthal} \&
  {Gould}}{1970}]{BG1970}
{Blumenthal} G.~R.,  {Gould} R.~J.,  1970, Reviews of Modern Physics, 42, 237

\bibitem[\protect\citeauthoryear{{Bohlin}, {Savage} \& {Drake}}{{Bohlin}
  et~al.}{1978}]{Boh}
{Bohlin} R.~C.,  {Savage} B.~D.,    {Drake} J.~F.,  1978, \apj, 224, 132

\bibitem[\protect\citeauthoryear{{Churazov}, {Gilfanov}, {Forman} \&
  {Jones}}{{Churazov} et~al.}{1996}]{churazov}
{Churazov} E.,  {Gilfanov} M.,  {Forman} W.,    {Jones} C.,  1996, \apj, 471,
  673

\bibitem[Condon et al.(1994)]{condon1994} Condon, J.~J., Cotton, W.~D., Greisen, E.~W., et al.\ 1994, Astronomical Data Analysis Software and Systems III, 61, 155

\bibitem[\protect\citeauthoryear{{De Becker}}{{De
  Becker}}{2007}]{debeckerreview}
{De Becker} M.,  2007, \aapr, 14, 171

\bibitem[\protect\citeauthoryear{{De Becker}}{{De
  Becker}}{2013}]{debecker2013}
{De Becker} M.,  2013, NewA, 25, 7

\bibitem[\protect\citeauthoryear{{De Becker}}{{De
  Becker}}{2015}]{debecker6604new}
{De Becker} M.,  2015, \mnras, 451, 1070

\bibitem[\protect\citeauthoryear{{De Becker} \& {Raucq}}{{De Becker} \&
  {Raucq}}{2013}]{catapacwb}
{De Becker} M.,  {Raucq} F.,  2013, \aap, 558, A28

\bibitem[de Boer et. al (2005)]{deBoer2005} de Boer, P.,  Kroese, D.~P., Mannor, S., \& Rubinstein, R.~Y.\ 2005, Annals of Operations Research, 134, 19

\bibitem[\protect\citeauthoryear{{del Valle} \& {Romero}}{{del Valle} \&
  {Romero}}{2012}]{ntbowshock}
{del Valle} M.~V.,  {Romero} G.~E.,  2012, \aap, 543, A56

\bibitem[\protect\citeauthoryear{{del Valle}, {Romero} \& {De Becker}}{{del
  Valle} et~al.}{2013}]{DRD2013}
{del Valle} M.~V.,  {Romero} G.~E.,    {De Becker} M.,  2013, \aap, 550, A112

\bibitem[del Valle et al.(2016)]{delvalle16} del Valle, M.~V., Lazarian, A., \& Santos-Lima, R.\ 2016, \mnras, 458, 1645 

\bibitem[\protect\citeauthoryear{{Dorman} \& {Arnaud}}{{Dorman} \&
  {Arnaud}}{2001}]{xspec2001}
{Dorman} B.,  {Arnaud} K.~A.,  2001, in {Harnden} Jr. F.~R.,  {Primini} F.~A.,
   {Payne} H.~E.,  eds, Astronomical Data Analysis Software and Systems X
  Vol.~238 of Astronomical Society of the Pacific Conference Series, {Redesign
  and Reimplementation of XSPEC}.
p.~415

\bibitem[\protect\citeauthoryear{{Drury}}{{Drury}}{1983}]{drury1983}
{Drury} L.~O'C., 1983, Rep. Prog. Phys., 46, 973

\bibitem[\protect\citeauthoryear{{Feldmeier}, {Puls} \&
  {Pauldrach}}{{Feldmeier} et~al.}{1997}]{FeldX}
{Feldmeier} A.,  {Puls} J.,    {Pauldrach} A.~W.~A.,  1997, \aap, 322, 878

\bibitem[\protect\citeauthoryear{{Fermi}}{{Fermi}}{1949}]{fermi}
{Fermi} E., 1949, Physical Review, 75, 1169

\bibitem[\protect\citeauthoryear{{H.E.S.S. Collaboration}, {Abdalla},
 {Abramowski} \& {et~al.}}{{H.E.S.S. Collaboration} et~al.}{2017}]{hessbs}
{H.E.S.S. Collaboration},  {Abdalla} H., {Abramowski} A., {et~al.} 2017, \aap, in press 

\bibitem[\protect\citeauthoryear{{Hohle}, {Neuh{\"a}user} \& {Schutz}}{{Hohle}
  et~al.}{2010}]{hns2010}
{Hohle} M.~M.,  {Neuh{\"a}user} R., {Schutz} B.~F.,  2010, Astronomische
  Nachrichten, 331, 349

\bibitem[\protect\citeauthoryear{{Le Bouquin}, {Sana}, {Gosset}, {De Becker},
  {Duvert} \& {et~al.}}{{Le Bouquin} et~al.}{2017}]{vlti2017}
{Le Bouquin} J.-B.,  {Sana} H.,  {Gosset} E.,  {De Becker} M.,  {Duvert} G.,
  {et~al.} 2017, \aap, 601, A34
  
\bibitem[\protect\citeauthoryear{{Leutenegger}, {Cohen}, {Zsarg\'o}, {Martell},
  {MacArthur} \& {et~al.}}{{Leutenegger} et~al.}{2010}]{Leut2010}
{Leutenegger} M.~A., {Cohen} D.~H., {Zsarg\'o} J., {Martell} E.~M.,  {MacArthur} J.~P., {et~al.} 2010, \aap, 719, 1767  

\bibitem[\protect\citeauthoryear{{Leonard} \& {Duncan}}{{Leonard} \&
  {Duncan}}{1990}]{leonardduncan}
{Leonard} P.~J.~T.,  {Duncan} M.~J.,  1990, \aj, 99, 608

\bibitem[Longair(2011)]{longairbook} Longair, M.~S.\ 2011, High Energy Astrophysics, by Malcolm S.~Longair, Cambridge, UK: Cambridge University Press, 2011

\bibitem[\protect\citeauthoryear{{L{\'o}pez-Santiago}, {Miceli}, {del Valle},
  {Romero}, {Bonito}, {Albacete-Colombo}, {Pereira}, {de Castro} \&
  {Damiani}}{{L{\'o}pez-Santiago} et~al.}{2012}]{aeaurigaexmm}
{L{\'o}pez-Santiago} J.,  {Miceli} M.,  {del Valle} M.~V.,  {Romero} G.~E.,
  {Bonito} R.,  {Albacete-Colombo} J.~F.,  {Pereira} V.,  {de Castro} E.,
  {Damiani} F.,  2012, \apjl, 757, L6

\bibitem[\protect\citeauthoryear{{Luhman} \& {Mamajek}}{{Luhman} \&
  {Mamajek}}{2012}]{LM2012}
{Luhman} K.~L.,  {Mamajek} E.~E.,  2012, \apj, 758, 31

\bibitem[\protect\citeauthoryear{{Lumb}, {Warwick}, {Page} \& {De Luca}}{{Lumb}
  et~al.}{2002}]{Lumb2002}
{Lumb} D.~H.,  {Warwick} R.~S.,  {Page} M.,  {De Luca} A.,  2002, \aap, 389,
  93

\bibitem[\protect\citeauthoryear{{Martins}, {Schaerer} \& {Hillier}}{{Martins}
  et~al.}{2005}]{martins}
{Martins} F.,  {Schaerer} D.,    {Hillier} D.~J.,  2005, \aap, 436, 1049

\bibitem[\protect\citeauthoryear{{Meilland}, {Delaa}, {Stee}, {Kanaan},
  {Millour}, {Mourard}, {Bonneau} \& {et~al.}}{{Meilland}
  et~al.}{2011}]{delsco2011}
{Meilland} A.,  {Delaa} O.,  {Stee} P.,  {Kanaan} S.,  {Millour} F.,  {Mourard}
  D.,  {Bonneau} D.,    {et~al.} 2011, \aap, 532, A80

\bibitem[\protect\citeauthoryear{{Mihalas} \& {Binney}}{{Mihalas} \&
  {Binney}}{1981}]{MB1981}
{Mihalas} D.,  {Binney} J.,  1981, {Galactic astronomy: Structure and
  kinematics, 2nd edition}

\bibitem[\protect\citeauthoryear{{Owocki}, {Sundqvist}, {Cohen} \&
  {Gayley}}{{Owocki} et~al.}{2013}]{owocki2013}
{Owocki} S.~P.,  {Sundqvist} J.~O.,  {Cohen} D.~H.,    {Gayley} K.~G.,  2013,
  \mnras, 429, 3379

\bibitem[\protect\citeauthoryear{{Peri}, {Benaglia}, {Brookes}, {Stevens} \&
  {Isequilla}}{{Peri} et~al.}{2012}]{ebossI}
{Peri} C.~S.,  {Benaglia} P.,  {Brookes} D.~P.,  {Stevens} I.~R.,
  {Isequilla} N.~L.,  2012, \aap, 538, A108

\bibitem[\protect\citeauthoryear{{Peri}, {Benaglia} \& {Isequilla}}{{Peri}
  et~al.}{2015}]{ebossII}
{Peri} C.~S.,  {Benaglia} P., {Isequilla} N.~L., 2015, \aap, 578, A45

\bibitem[\protect\citeauthoryear{{Puls}, {Vink} \& {Najarro}}{{Puls}
  et~al.}{2015}]{puls2008}
{Puls} J.,  {Vink} J.~S., {Najarro} F., 2008, \aapr, 16, 209

\bibitem[\protect\citeauthoryear{{Pittard} \& {Dougherty}}{{Pittard} \&
  {Dougherty}}{2006}]{PD140art}
{Pittard} J.~M.,  {Dougherty} S.~M.,  2006, \mnras, 372, 801

\bibitem[\protect\citeauthoryear{{Romero}}{{Romero}}{2004}]{romerosnr}
{Romero} G.~E.,  2004, in {K.~S.~Cheng \& G.~E.~Romero} ed., Cosmic Gamma-Ray
  Sources Vol.~304 of Astrophysics and Space Science Library, {Gamma-ray
  emission from supernova remnants}.
p.~127

\bibitem[\protect\citeauthoryear{{Rosen}, {Webb}, {Watson}, {Ballet}, {Barret},
  {Braito}, {Carrera} \& {et~al.}}{{Rosen} et~al.}{2016}]{XMMSS3}
{Rosen} S.~R.,  {Webb} N.~A.,  {Watson} M.~G.,  {Ballet} J.,  {Barret} D.,
  {Braito} V.,  {Carrera} F.~J.,    {et~al.} 2016, \aap, 590, A1

\bibitem[\protect\citeauthoryear{{Schulz}, {Ackermann}, {Buehler}, {Mayer} \&
  {Klepser}}{{Schulz} et~al.}{2014}]{schulzbsr}
{Schulz} A.,  {Ackermann} M.,  {Buehler} R.,  {Mayer} M.,    {Klepser} S.,
  2014, \aap, 565, A95

\bibitem[\protect\citeauthoryear{{Skrutskie}, {Cutri}, {Stiening}, {Weinberg},
  {Schneider}, {Carpenter}, {Beichman} \& {et~al.}}{{Skrutskie}
  et~al.}{2006}]{2mass}
{Skrutskie} M.~F.,  {Cutri} R.~M.,  {Stiening} R.,  {Weinberg} M.~D.,
  {Schneider} S.,  {Carpenter} J.~M.,  {Beichman} C.,    {et~al.} 2006, \aj,
  131, 1163

\bibitem[\protect\citeauthoryear{{Tango}, {Davis}, {Jacob}, {Mendez}, {North},
  {O'Byrne}, {Seneta} \& {Tuthill}}{{Tango} et~al.}{2009}]{delsco2009}
{Tango} W.~J.,  {Davis} J.,  {Jacob} A.~P.,  {Mendez} A.,  {North} J.~R.,
  {O'Byrne} J.~W.,  {Seneta} E.~B.,    {Tuthill} P.~G.,  2009, \mnras, 396, 842

\bibitem[\protect\citeauthoryear{{Toal{\'a}}, {Oskinova},
  {Gonz{\'a}lez-Gal{\'a}n}, {Guerrero}, {Ignace} \& {Pohl}}{{Toal{\'a}}
  et~al.}{2016}]{toala2016}
{Toal{\'a}} J.~A.,  {Oskinova} L.~M.,  {Gonz{\'a}lez-Gal{\'a}n} A.,  {Guerrero}
  M.~A.,  {Ignace} R.,    {Pohl} M.,  2016, \apj, 821, 79
  
\bibitem[\protect\citeauthoryear{{Toal{\'a}}, {Oskinova},
  \& {Ignace}}{{Toal{\'a}} et~al.}{2017}]{toala2017}
{Toal{\'a}} J.~A.,  {Oskinova} L.~M., {Ignace} R., 2017, \apj, 838, L19

\bibitem[van Buren \& McCray(1988)]{vanburen1988} van Buren, D., \& McCray, R.\ 1988, \apjl, 329, L93 

\bibitem[\protect\citeauthoryear{{Vink}}{{Vink}}{2013}]{VinkSNRutrecht}
{Vink} J.,  2013, in {Lamers} H.,  {Pugliese} V.,  eds, 370 years of astronomy
  in Utrecht Astronomical Society of the Pacific Conference Series, {Supernova
  Remnants as the Sources of Galactic Cosmic Rays}

\bibitem[\protect\citeauthoryear{{Zwicky}}{{Zwicky}}{1957}]{zwicky}
{Zwicky} F., 1957, {Morphological astronomy}.
Berlin: Springer, 1957

\end{thebibliography}








\bsp	
\label{lastpage}
\end{document}